\documentclass[]{aa}  

\usepackage{graphicx}

\usepackage{txfonts}

\usepackage[breaklinks=true, colorlinks=true,linkcolor=blue, citecolor=blue]{hyperref}

\usepackage{layouts}
\usepackage{multirow}

\newcommand{\grad}{{\rm grad}\,}

\begin{document}

   \title{A new framework of multidimensional pulsating stellar envelopes I.}

   \subtitle{ Properties of turbulent convection in static RR Lyrae envelope models with SPHERLS}

   \author{Gábor B. Kovács
          \inst{1,2,3}
          ,
          R. Szabó\inst{1,2}
          \and
          J. Nuspl\inst{1}
          }

   \institute{HUN-REN Research Centre for Astronomy and Earth Sciences, Konkoly Observatory, MTA Centre of Excellence, 
              Konkoly Thege Miklós út 15-17., H-1121 Budapest, Hungary\\
              \email{kovacs.gabor@csfk.org}
            \and
             Eötvös Loránd University, Institute of Physics and Astronomy, Pázmány Péter sétány 1/a,  H-1117 Budapest, Hungary
             \and
             ELTE E\"{o}tv\"{o}s Lor\'{a}nd University, Gothard Astrophysical Observatory, Szombathely, Szent Imre h. u. 112., H-9700, Hungary
             }

   \date{Received 19 Sept 2024; accepted 12 Jun 2025}

  \abstract
   {The one-dimensional treatment of turbulent convection had large successes 
   until the early 2000s. However, the recent abundance and precision of observational data 
   shows that this problem is far from solved.  
   Even so, ongoing theoretical debates about proper one-equation-based treatment of convection and new results show that it has various other theoretical difficulties as well. A more modern approach should be developed by using multidimensional models.}
   {We established a new theoretical framework for comparison between one-dimensional and multidimensional convection models by mapping the two-dimensional structure of the convective zone and optimizing the modeling parameters of the SPHERLS code.}
   {We  constructed a series of  static envelope models for the same RR Lyrae stars, but with different horizontal sizes and resolutions. We then used a series of statistical methods to quantify the sizes of convective eddies, map the energy cascade, and describe the different structural parts of the convective zone. These include integral length scales, Fourier series, and the determination of the convective flux through horizontal averaging.}
   {The structure of the convective zone depends significantly on the model size below an angular size of $9^\circ$. Models of at least this size are more consistent, and the horizontal resolution of earlier studies is adequate to describe the granulation pattern in the large eddy simulation approach. In quasi-static RR Lyrae stars, the convective zone consists of two distinct dynamically unstable regions that are loosely connected. Approximately half of the convective flux is supplied by the transport of ionization energy in the partial hydrogen ionization zone.}
   {The 2D models presented in this work with the described size and resolution parameters can be used for comparison against 1D models. The structure of the convective zone urges reconsideration of some recent approaches to describe the convective flux currently used in radial stellar pulsation codes, which will be addressed in a separate paper. }


   \maketitle

\section{Introduction}

Convection is a buoyancy-driven transport process \citep{conv-book}, formed when the temperature gradient is greater than the adiabatic temperature gradient in stars.\footnote{Thermodynamically,  an inhomogeneity in intensive parameters (e.g., chemical potential gradients) can force convective currents of extensive quantities.}
 Hence, small fluctuations in density cause mass elements to rise and sink nearly adiabatically, creating a 3D dimensional flow that is random, but has quasi-stationary structures called Rayleigh-Bénard cells. The flow inside these cells carries energy from the hotter bottom of the cell to the cooler top where the extra energy is released to the surroundings through heat diffusion and viscosity. This way, it transports energy, decreasing the temperature gradient of the system. In a closed system, this can sufficiently decrease this gradient to a level where the convection can stop, and the kinetic energy of the cells would dissipate. However, in an open system, for example a stellar envelope,  the temperature difference between the two sides of the convectively unstable region does not disappear, so buoyancy maintains the convective cells. This phenomenological picture was formulated in many different versions of mixing length theories (MLT).

However, in stellar envelopes (also in  stellar cores and other astrophysical fluids), the material viscosity of the fluid is very low. This causes the structure of Rayleigh-Bénard cells to decompose into smaller eddies, which  also decompose, and so on, until they reach small-scale motions that are effectively braked  by viscous forces. This cascade of eddies shows an inherently unpredictable chaotic flow structure that is called turbulent flow \citep{Pope-konyv}, and its appearance can be characterized by a dimensionless number, the Reynolds number, determined by the fluid physical parameters.\footnote{The turbulence arises from instability and is closely connected to chaos observed in nonlinear dynamical systems; we cannot follow the path of a small mass element, although it is one of the basic assumptions in fluid mechanics. For example, the particles of a small dye drop will disperse to an attractor with fractal geometry in the phase space.}
This number, by definition, is the ratio of the inertial to viscous forces, $Re={uL}/{\nu}$, with a characteristic velocity, $u$, the characteristic length scale, $L$, and the kinetic viscosity, $\nu$, of the flow. If $Re \gg 1$, a flow becomes turbulent (e.g., its value is  $Re \sim 10^{12}$ \citep{kupka} in the solar envelope). Hence, convection in stars is always turbulent, but we emphasize that the two phenomena are different, and one can occur without the other, outside the scope of stellar astrophysics.

These are genuinely 3D phenomena, and their description in 1D pulsation models can only be achieved through approximations. First, one needs to use averaging on the 3D Navier-Stokes equations to derive horizontal mean quantities that are changing with the pulsations \citep[such a derivation is presented by][]{Kuhfuss1986}. These are called ensemble averages (\citealt{Pope-konyv} and for usage in other astrophysics problems, see \citealt{kupka}). Unfortunately, the method always includes  unspecified terms; therefore, the 1D equations are not closed. This closure problem can be solved with simplifications, and the level of complexity depends on the number and quality of the used approximations. In pulsation theory, these include the Boussinesq approximation meaning neglecting density fluctuations except for buoyancy \citep[first used for pulsations by][]{Gough1969}, the neglect of acoustic modes, the eddy viscosity hypothesis \citep{Kuhfuss1986} simplifying pressure terms, and the gradient-diffusion (or down-gradient) hypothesis \citep{Stellingwerf1982a}, which relates the unsolved transport terms to the gradient of the mean-field. We refer to \citet{kupka} and \citet{Houdek2015} for background material.

The other solution for the problem would be 3D modeling, directly solving the Navier-Stokes equations. The direct numerical simulation (DNS) is still too computationally expensive on modern computers as one would need to resolve scales of the flow from   scales of  several million kilometers to the scales of meters \citep{kupka}. Instead, one can use the so-called large eddy simulation (LES) method, pioneered by \citet{Smagorinsky1963}, \citet{Lilly1969},  and \citet{Deardorff1970}. The main idea is to resolve the large-scale, mainly anisotropic (in our case, buoyancy-driven convection) motions and model only those on this scale. These  modeling results are similar to measurements for a given resolution, or more likely, a filtered view of good resolution data. Regardless of the details, the filtering is similar to the ensemble averaging of the 1D models, although it has different properties \citep{Pope-konyv}. The method itself has been widely used in engineering and meteorology since the sixties \citep{Pope-konyv}; while it has had many successes in the simulation of solar granulation and solar physics \citep{Nordlund2009}, it is used for modeling A type stellar envelopes \citep{Kupka2023b}, studies of stellar cores \citep[see, e.g.,][]{Rizutti2023,Georgy2024}, AGB stars \citep[e.g.,][]{Freytag2012}, and stellar convection in general \citep{Meakin2007,Arnett2009,Viallet2013}.\footnote{Strictly speaking the type of models that \citet{Arnett2009} use, are the so called iLES models, or implicit LES models. In this type of modeling the conservative numerical method is used as filtering, hence there is no subgrid-scale modeling.} \citet{kupka} gives an overview of the history of LES in astrophysics.\footnote{We refer to the excellent review of \citet{kupka}, which gives a more detailed overview of the numerical approaches.}
However, we  emphasize that expectations  that the 3D simulations will solve all the problems via brute force are misleading.

There are two approaches that are widely used in the aforementioned studies: the so-called box-in-a-star and the star-in-a-box method. In the former, one defines a simulation box (horizontally with periodic boundary conditions) inside the star \citep[see, e.g.,][]{Stein2001,Muthsam2010} because the length scales of the convective motions are not comparable to the size of the star (this is the case in main-sequence stars and classical pulsators). The latter case is used for red and asymptotic giant branch stars, where the convective length scales are comparable to the stellar radius \citep[e.g.,][]{Freytag2012}.

In the case of radial stellar pulsations of classical variables, a further problem occurs, namely the large amplitude radial oscillation of the outer layers, where the convection zone resides. This means that any codes need to follow the movement of the stellar layers; meanwhile, convection needs to remain  resolved. If the former condition is not met, one cannot reach full-amplitude pulsation with the model, as the important partially ionization zones moves out from the grid \citep{Deupree1977a}. The first attempt to reproduce 2D pulsation and convection together was done by \citet{Deupree1977a}, successfully reproducing the red edge of the instability strip \citep{Deupree1977b,Deupree1980,Deupree1985}. Recently \citet{Mundprecht2013} implemented an adaptive method into the {\tt ANTARES} code \citep{Muthsam2010} which was used to simulate Cepheids \citep{Mundprecht2015}. \citet{Vasilyev2018} used the {\tt COBOLD5} software \citep[][and references therein]{Vasilyev2017} to simulate a 2D Cepheid atmosphere as an input for spectral synthesis models. \citet{SPHERLS1,SPHERLSII} developed the SPHERLS code based on the ideas of \citet{Deupree1977a} and successfully reached full-amplitude RR Lyrae models \citep{SPHERLSIII,SPHERLS4}, and we are using this code in our current study.

In this paper in our series, we optimize the SPHERLS code input parameters (resolution, model size) and determine the structure of the convective zone through a series of 2D quasi-static simulations to establish a theoretical ground for the comparison with 1D. For this, we use statistical methods and dimensionless numbers largely used in turbulence theory \citep{Pope-konyv}, and define some new to follow the effects of the numerics.  For example, we are using nondimensional numbers, such as the Prandtl number, that characterize the fluid and can be compared to numbers of real stellar flows derived from material science \citep{kupka}, while we are using numbers that characterize the flow, for example the Reynolds number, which is also dependent on the simulation setup.

In addition to the statistical quantities, we define other quantities by describing the size and structure of the convective zone, as  these are comparable directly to 1D models. Only the consistency of these and statistical quantities can ensure that a comparison with 1D calculations, and the conclusions  derived from it, would be meaningful. 

In our previous papers \citep[][and references therein]{KovacsGB2023,KovacsGB2024}, we confronted two 1D models (calculated with the Budapest-Florida and MESA-RSP codes) of RR Lyrae stars in M3 globular cluster with observed radial velocities, calibrated their free parameters, and analyzed the processes occurring in them during pulsation. This work focuses on technical questions of 2D modeling of these stars, and the following papers will discuss in detail the results provided by the comparison of 1D codes and the multidimensional SPHERLS in 2D for static and pulsating states.

The structure of this paper is the following. First, we introduce the main concepts behind the SPHERLS code alongside with the stellar parameters of the input models in Section \ref{sec:models}. Then we define our analytical methods in Section \ref{sec:methods}: The modeling procedure (Sec. \ref{sec:mod_proc}), the statistical methods that are describing turbulent flows (Sec. \ref{sec:stat_quant}), and the various dimensionless quantities related to turbulence, numerics, and 1D treatment (Sec. \ref{sec:dimless}. This is followed by the description of our results. First, we present the numerical costs of our calculations in Sec. \ref{sec:numcost}, then we describe the vertical structure of the convective zone in the studied quasi-static RR Lyrae models in Section \ref{sec:conv_struct}. We studied the different components of the convective flux, which is presented in Section \ref{sec:components}. The effects of resolution and horizontal model size on the structure of the velocity field, on the size of the eddies and on the behavior of the energy cascade is given in Sections \ref{sec:veolicity}, \ref{sec:lengthscales}, and \ref{sec:energy}, respectively. Our results and their consequences are discussed in Section \ref{sec:discussion}, and our conclusions and a summary are given in Section \ref{sec:conclusion}.

\section{The multidimensional SPHERLS model}
\label{sec:models}

SPHERLS is an open-source multidimensional numerical hydrocode\footnote{ Available on GitHub under MIT Licence: \url{https://github.com/cgeroux/SPHERLS}}  that was developed by \citet{SPHERLS1,SPHERLSII} for modeling stellar pulsation using LES principles.
This code is based on a hybrid Lagrangian-Eulerian scheme, which makes it possible to follow the long-term and short-term changes in radially pulsating variable stars while maintaining a good enough resolution of the ionization regions. We repeat the governing equations in Appendix~\ref{appendix:SPHERLS} to match our notations.

 In the field of classical stellar pulsations, the main focus is on the optically thick stellar envelope, in which the partial ionization zones that drive the pulsations reside. This means that the atmosphere itself is used as a boundary condition and radiation flux is described through  the diffusion approximation. SPHERLS has limited capabilities regarding horizontal resolution (see Sec.~\ref{sec:numcost}), even with simple numeric methods (see Appendix \ref{appendix:numerics} and \ref{appendix:tests}). With horizontal resolution being magnitudes lower than current hydro-codes used in solar convection modeling, and following the same principles as 1D pulsation codes, this modeling method can be seen as an extended version of the latter, but without the Mixing Length Theory and other 1D approximations. So this code is suitable to study stellar convection in pulsating stars, keeping in mind the aforementioned limitations we have.

The three primary numerical concepts crucial to our study are the box-in-a-star nature of SPHERLS, the utilization of the subgrid scale viscosity model  and the effects of numerical damping.

The box-in-a-star approach of SPHERLS means that we are modeling a slice of an angular shell of the star.  \citet{SPHERLS4} have shown that 2D calculations are adequate to describe RR Lyrae variables. Therefore, we   calculated 2D wedges, keeping in mind that in 2D the average lifetime of a convective structure can be longer \citep{Kupka2020} and the cascade behavior is different than in the 3D case \citep{Kupka2009_LNP}. On the horizontal boundaries, we use periodic boundary conditions; the top boundary of the star is determined at the surface with a constant mass closing shell and pressure \citep{SPHERLSII}. The bottom boundary is taken at 2~million~K temperature with an incompressible boundary. These vertical boundaries are the same which are used in radial pulsation models  in general \citep[see, e.g.,][]{Paxton2019}, and the stellar parameters define the vertical extent of the model. In contrast, the horizontal extent is always chosen to an arbitrary angular width that we call $\theta_{\rm width}$.  This horizontal extent is defined by the number of horizontal cells, $N_\theta$, and the horizontal size of a single cell, $\Delta\theta$, as $\theta_{\rm width}=N_\theta \Delta\theta$. The choice of  these values  is not straightforward, as one has to choose  them to have a $\theta_{\rm width}$  large enough to capture the large-scale convective processes but  with small enough $\Delta\theta$  to maintain the necessary resolution.

Another essential concept is the subfilter or subgrid scale (SGS) model. Since the simulation only solves equations on the so-called resolved scales, the effects on smaller scales than the grid scales have to be modeled somehow. Such an effect is, for instance, the turbulent cascade itself: the kinetic energy of the resolved scales is transferred to these SGS scales, and without proper treatment,  the simulations fail. The SGS model usually defines the filter function of the LES equations, as well. The length scale of this function $\Delta_{\rm filt}$ is coupled to the resolution where the LES method gives information \citep{Magnient2007}. The length scale regarding the dissipation of energy ($\Delta_{\rm diss}$) is related to $\Delta_{\rm filt}$ by the applied SGS model. This, in practice, means that an LES calculation shows the dissipation of kinetic energy at a scale of a different length than the actual flow. In the case of the SPHERLS, chose $\Delta_{\rm filt}=l$ where $l$ is the length scale of the grid (see Appendix~\ref{appendix:SPHERLS}). 

SPHERLS uses the Smagorinsky-model \citep{Smagorinsky1963}, which connects the viscous effects of SGS turbulence to the velocity stress tensor, similar to the eddy viscosity models in 1D; this also assumes incompressible turbulence.  Although we perform compressible hydrodynamics, one can assume that on the subgrid scales, the fluid is incompressible. This assumption is valid if the related turbulent velocities are subsonic \citep{LES_konyv} even in the case when the background radial pulsation  has velocities above the speed of sound. 

 Lastly, numerical damping is a serious issue, especially in low-resolution models. This damping is caused by the numerical scheme of every hydrocode, and its main source is the truncation error of the spatial derivatives. This effect also can be stronger than the effects of the SGS models. We discuss this in detail in Appendix~\ref{appendix:numerics}. If this damping is too strong, shockwaves cannot be modeled. This could be an issue in the modeling of the ionization front, for which case we performed additional numerical tests next to the original ones of \citet{SPHERLS1}, which we present in Appendix~\ref{appendix:tests}.   

Similar to \cite{SPHERLSII}, we use tabulated opacity and equation of state values by OPAL \citep{OPAL1994} and the low temperature \citet{Alexander1994} tables.

\begin{table}[ht]
    \centering
    \caption{Input parameters of the model series.}
    \label{tab:stellar_properties}
    \begin{tabular}{cccccc}
    \hline
    \hline
         Series & $M$ & $L$ & $T$ & $X$ & $Z$  \\
         \hline
         v036 & $0.548\, M_\odot$ & $44.2\, L_\odot$&  $6678\,{\rm K}$&$0.77$&  $0.0005$ \\
         v046 & $0.571\, M_\odot$ & $43.8\,L_\odot$ &$6394\,{\rm K}$ & $0.73$& $0.0006$
    \end{tabular}
    \tablefoot{The columns are (from left to right) the stellar mass, luminosity, effective temperature, hydrogen mass fraction, and metallicity. The values are from \citet{KovacsGB2023}}
\end{table}

We ran a series of 2D models with two different sets of stellar parameters. These represent the RRab stars v036 and v046 in the M3 globular cluster, which were used for 1D model calibration by \citet{KovacsGB2023}. We present the input parameters in Table~\ref{tab:stellar_properties}.

 \cite{SPHERLSII} used originally an angular size of $\theta_{\rm width}=6^\circ$ and horizontal cell number of $N_\theta=20$, claiming that it is enough for one upward and downward moving convective cells to form. We refer to this model as the reference model throughout our paper. We chose a number of different angular widths for the models, covering the range from $4^\circ$ to $45^\circ$. The number of horizontal cells chosen is between 20 and 60; this way we  ran bad-resolution and small-sized models to test the effects of these arbitrary parameters.

The vertical resolution of our models is based on the 1D resolutions of pulsating stars \citep{Paxton2019}: we have a total $N_r$ number of radial Lagrangian shells from which $N_u$  is uniformly dispatched above the so-called anchor temperature, which is chosen to be $T_{\rm anchor} = 11000$ K in the partial hydrogen ionization region. Below this temperature, the shell sizes are increased according to a geometrical series. We ran 3 different vertical resolutions: $N_r=150$, $N_u=50$ is our baseline resolution ($N_r=152$ in the case of v046 models), we have a few models calculated with $N_r=122$, $N_u=25$, and a high-resolution model with $N_r=200$, $N_u=75$, $T_\text{anchor}=50000$ K. Meanwhile, due to the very small cell sizes and the otherwise damped horizontal motions, SPHERLS uses 1D cells at the bottom of the model. We have 25 1D cells in the bottom of our baseline of low vertical resolution models, and 50 cells in the larger vertical resolution models.  The different model dimensions are presented in Table~\ref{tab:model_properties}.

 Due to this setup, the aspect ratio of the grid cells are changing throughout the stellar envelope: since the horizontal size is $\sim r$, we have a finer horizontal grid deeper in the star, while in the region above $T_\text{anchor}$ the vertical resolution is finer, with aspect ratios $\Delta r \colon r\Delta\theta \sim 1\colon 40$ or larger. This directly affects the numerical damping and SGS viscosity: in some regions we have stronger horizontal damping then in other regions. We present aspect ratios of our models in Table~\ref{tab:appendix1}, in Appendix~\ref{appendix:numbers}.

\begin{table*}[ht!]
    \centering
    \caption{
    General properties and dimensionless quantities of the models}
    \label{tab:model_properties}
    \begin{tabular}{lccccccccccc}
    \hline
    \hline
        Name & $\theta_{\rm width}$ & $N_r$ & $N_\theta$ & $Re_{\rm ch}$ & $Pr_{\rm ch}$ &$Pr_\text{HI}$ & $Pe_\text{HI}$ & $Nu$ & $q_{\rm SGS}$ & $q_{\rm num}$& $q_{\rm diff}$\\
        \hline
        
\verb|v036_4_150x20| & $  4$ & $150$ & $ 20$ & $28.4$ & $0.13$ & $0.04$ & $0.82$ & $1.12$ &  $1.19\times 10^{-3}$ & $9.20\times 10^{1}$ & $4.813\times 10^{3}$ \\
\verb|v036_4_150x30| & $  4$ & $150$ & $ 30$ & $42.6$ & $0.08$ & $0.02$ & $0.55$ & $1.09$ &  $7.66\times 10^{-4}$ & $1.34\times 10^{2}$ & $4.799\times 10^{3}$ \\
\hline
\verb|v036_6_150x20| & $  6$ & $150$ & $ 20$ & $19.0$ & $0.19$ & $0.04$ & $0.57$ & $1.07$ &  $1.07\times 10^{-3}$ & $1.22\times 10^{2}$ & $2.967\times 10^{3}$ \\
\verb|v036_6_150x30| & $  6$ & $150$ & $ 30$ & $28.4$ & $0.13$ & $0.04$ & $0.86$ & $1.08$ &  $8.56\times 10^{-4}$ & $1.28\times 10^{2}$ & $5.152\times 10^{3}$ \\
\verb|v036_6_150x40| & $  6$ & $150$ & $ 40$ & $37.8$ & $0.09$ & $0.01$ & $0.38$ & $1.06$ &  $5.21\times 10^{-4}$ & $1.35\times 10^{2}$ & $2.205\times 10^{3}$ \\
\verb|v036_6_150x50| & $  6$ & $150$ & $ 50$ & $47.5$ & $0.08$ & $0.02$ & $0.52$ & $1.07$ &  $5.32\times 10^{-4}$ & $1.27\times 10^{2}$ & $4.129\times 10^{3}$ \\
\verb|v036_6_201x20| & $  6$ & $201$ & $ 20$ & $18.4$ & $0.19$ & $0.04$ & $0.63$ & $1.10$ &  $3.56\times 10^{-4}$ & $2.35\times 10^{2}$ & $6.873\times 10^{2}$ \\
\hline
\verb|v036_8_150x20| & $  8$ & $150$ & $ 20$ & $14.2$ & $0.25$ & $0.04$ & $0.51$ & $1.04$ &  $1.09\times 10^{-3}$ & $1.15\times 10^{2}$ & $2.155\times 10^{3}$ \\
\verb|v036_8_150x30| & $  8$ & $150$ & $ 30$ & $21.3$ & $0.17$ & $0.04$ & $0.68$ & $1.06$ &  $8.07\times 10^{-4}$ & $1.14\times 10^{2}$ & $4.160\times 10^{3}$ \\
\verb|v036_8_150x40| & $  8$ & $150$ & $ 40$ & $28.5$ & $0.13$ & $0.03$ & $0.57$ & $1.06$ &  $6.80\times 10^{-4}$ & $1.18\times 10^{2}$ & $2.338\times 10^{3}$ \\
\hline
\verb|v036_9_150x30| & $  9$ & $150$ & $ 30$ & $18.9$ & $0.19$ & $0.03$ & $0.55$ & $1.06$ &  $8.26\times 10^{-4}$ & $1.14\times 10^{2}$ & $2.585\times 10^{3}$ \\
\hline
\verb|v036_10_150x20| & $ 10$ & $150$ & $ 20$ & $11.4$ & $0.32$ & $0.03$ & $1.15$ & $1.04$ &  $1.45\times 10^{-3}$ & $1.04\times 10^{2}$ & $5.860\times 10^{3}$ \\
\verb|v036_10_150x30| & $ 10$ & $150$ & $ 30$ & $17.1$ & $0.21$ & $0.06$ & $1.02$ & $1.07$ &  $1.19\times 10^{-3}$ & $1.09\times 10^{2}$ & $4.348\times 10^{3}$ \\
\hline
\verb|v036_12_122x60| & $ 12$ & $122$ & $ 60$ & $28.3$ & $0.13$ & $0.02$ & $0.52$ & $1.05$ &  $8.06\times 10^{-4}$ & $7.56\times 10^{1}$ & $6.559\times 10^{2}$ \\
\verb|v036_12_150x20| & $ 12$ & $150$ & $ 20$ & $9.5$ & $0.38$ & $0.06$ & $0.94$ & $1.03$ &  $1.43\times 10^{-3}$ & $1.54\times 10^{2}$ & $4.282\times 10^{3}$ \\
\verb|v036_12_150x30| & $ 12$ & $150$ & $ 30$ & $14.2$ & $0.25$ & $0.05$ & $1.02$ & $1.02$ &  $1.15\times 10^{-3}$ & $1.29\times 10^{2}$ & $6.612\times 10^{3}$ \\
\verb|v036_12_150x40| & $ 12$ & $150$ & $ 40$ & $18.9$ & $0.19$ & $0.03$ & $0.58$ & $1.05$ &  $7.44\times 10^{-4}$ & $1.31\times 10^{2}$ & $1.390\times 10^{3}$ \\
\hline
\verb|v036_15_150x50|\tablefootmark{$\star$} & $ 15$ & $150$ & $ 50$ & $18.8$ & $0.19$ & $0.03$ & $0.61$ & $1.03$ &  $7.65\times 10^{-4}$ & $1.49\times 10^{2}$ & $1.950\times 10^{3}$ \\
\hline
\verb|v036_18_150x60|\tablefootmark{$\star$} & $ 18$ & $150$ & $ 60$ & $19.0$ & $0.19$ & $0.04$ & $0.73$ & $1.06$ &  $7.25\times 10^{-4}$ & $1.01\times 10^{2}$ & $2.053\times 10^{3}$ \\
\hline
\verb|v036_45_122x40| & $ 45$ & $122$ & $ 40$ & $5.0$ & $0.71$ & $0.03$ & $0.19$ & $1.02$ &  $3.25\times 10^{-3}$ & $8.35\times 10^{1}$ & $1.823\times 10^{2}$ \\
\verb|v036_45_122x60| & $ 45$ & $122$ & $ 60$ & $7.5$ & $0.47$ & $0.02$ & $0.24$ & $1.03$ &  $2.28\times 10^{-3}$ & $8.89\times 10^{1}$ & $2.742\times 10^{2}$ \\
\verb|v036_45_150x20| & $ 45$ & $150$ & $ 20$ & $2.4$ & $1.42$ & $0.03$ & $0.13$ & $1.01$ &  $5.99\times 10^{-3}$ & $1.69\times 10^{2}$ & $1.710\times 10^{2}$ \\
\hline
\verb|v046_6_152x20| & $  6$ & $152$ & $ 20$ & $19.8$ & $0.21$ & $0.06$ & $1.06$ & $1.28$ &  $1.03\times 10^{-3}$ & $3.11\times 10^{2}$ & $9.639\times 10^{3}$ \\
\verb|v046_6_152x30| & $  6$ & $152$ & $ 30$ & $29.2$ & $0.14$ & $0.04$ & $0.92$ & $1.11$ &  $7.32\times 10^{-4}$ & $1.85\times 10^{2}$ & $3.239\times 10^{3}$ \\
\verb|v046_12_152x40| & $ 12$ & $152$ & $ 40$ & $19.6$ & $0.21$ & $0.06$ & $1.17$ & $1.10$ &  $9.38\times 10^{-4}$ & $1.42\times 10^{2}$ & $2.859\times 10^{3}$ \\
            \end{tabular}
    \tablefoot{The panels are separated according to the angular size and physical parameters of the models. The columns are (from left to right) the  model name,  angular extent  ($\theta_{\rm width}$),  number of radial shells ($N_r$),  number of horizontal cells ($N_\theta$),   characteristic Reynolds number ($Re_{\rm ch}$)  defined by eq. (\ref{eq:Re_ch}),  characteristic Prandtl number ($Pr^{\rm ch}$) defined by eq.~(\ref{eq:Pr_ch}), effective Prandtl number  in the HI region ($Pr_{\rm HI}$),  Péclet number in the  HI region ($Pe_{\rm HI}$),  maximum Nusselt number ($Nu$),  median ratio of resolved to unresolved (subgrid scale) kinetic energy ($q_{\rm SGS}$), the median ratio of numerical to SGS viscosity ($q_{\rm num}$) and the median ratio of numerical to SGS energy diffusion ($q_{\rm diff}$). \\
     \tablefoottext{$\star$}{This model did not reach $\tau_{\rm relax}$ simulation time, due to the large numerical cost; therefore, it is not reliable. It is only included here to show the trends. Statistical evaluation was performed over an interval of 12--18 days.}
    }
\end{table*}

\section{Methods}
\label{sec:methods}

We restricted our investigation to quasi-static envelopes before pulsation is triggered in them. In these envelopes, convection, and turbulence already develop spontaneously, which can thus be studied purely in themselves and their characterization is our aim in this study.

\subsection{The modeling procedure}
\label{sec:mod_proc}
\begin{figure*}[tp]

\begin{center}
    \includegraphics[width=\textwidth]{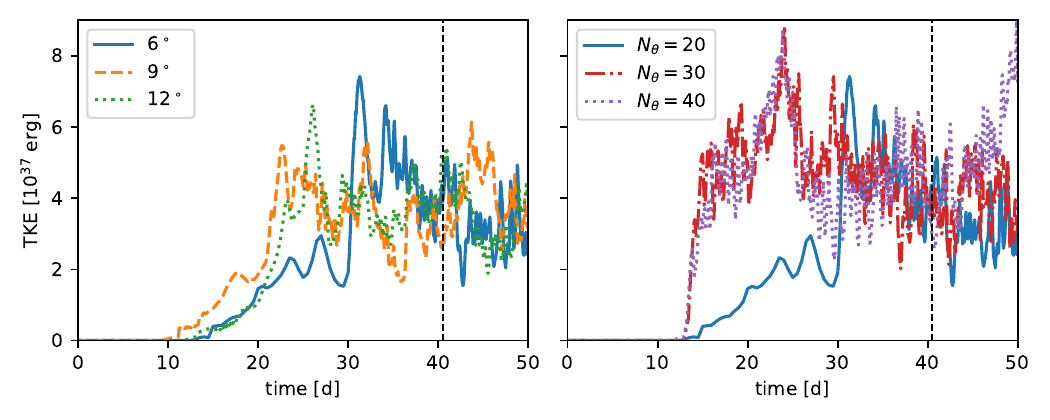}
    \caption{Time evolution of the total turbulent kinetic energy (TKE). Left panel: Comparison of the TKE evolution for different opening angles with resolution of $\Delta\theta = 0.3^\circ$. The models are {\tt v036\_6\_150x20 } (blue solid line), {\tt v036\_9\_150x30 } (yellow dashed line), and {\tt v036\_12\_150x40 } (green dotted line).  Right panel: Same as left panel, but with the same opening angle ($\theta_{\rm with}=6^\circ$) and different resolutions: {\tt v036\_6\_150x20 } (blue solid line), {\tt v036\_6\_150x30 }  (red dash-dotted line), and {\tt v036\_6\_150x40 } (purple dotted line). The vertical dashed line denotes the end of $\tau_{\rm relax}$ and the start of $\tau_{\rm stat}$.}
    \label{fig:TKE-evol}
\end{center}
\end{figure*}

The discussed models are quasi-static which means that we started the hydrodynamic calculations from the  zero velocity state in contrast with the method used in pulsation physics \citep[see, e.g.,][]{lengyel,SPHERLSII,KovacsGB2023}. 
As these are pulsationally unstable envelopes, which means that a small perturbation (caused by either the convective motions or the numerical errors) can excite them, the models could not be run for too long; some models (e.g., {\tt v036\_6\_150x40}) showed small amplitude but detectable pulsations at the end of the simulations. As the main focus was on the LES methodology of the SPHERLS code and not on the pulsation for the current study, we   adjusted the simulation times accordingly.

All models were started from a fully radiative static initial structure \citep[following the procedure of][]{SPHERLSII}, and convection is ignited automatically by the numerical truncation errors in quantities of the models. Since the convective regions in RR Lyrae variables are relatively shallow, the convective turnover timescale ($\tau_{\rm conv}$) is 2--5 days. To reach a steady state of the convection, we ran the models for 40 days of simulation time, which sets the relaxation timescale \citep{kupka} $\tau_{\rm rel} \sim 8-10\ \tau_{\rm conv}$. The statistical analysis of the models started after 40.5 days, and the statistical timescale was chosen to be 9 days ($\tau_{\rm stat} \sim 2-3\ \tau_{\rm conv}$); therefore, the  statistical data were collected over the interval of  $\sim(40-50)$ days. The pulsation amplitude starts to increase only after that point, and was where we stopped the simulations.

In Fig.~\ref{fig:TKE-evol}. we present the total turbulent kinetic energy (hereafter TKE) evolution of six different realizations of the v036 model. The differences are in the angular width (left panel) and angular resolution (right panel). The start of the statistical analysis is denoted by the vertical dashed line. The total turbulent kinetic energy is the kinetic energy of the shells projected onto the entirety of the star,
\begin{equation}
    E_{\rm TKE} =\mathcal{S^\star} \frac{1}{2} \int\limits_{R_0}^R \int\limits_{\theta_0}^{\theta_0+\theta_{\rm width}} \rho (r,\theta)\, |\boldsymbol U(r,\theta) -\boldsymbol U_0(r)|^2 r^2 \sin\theta {\rm d}\theta {\rm d}r  \text{,}
\end{equation}
where $\rho$ is the density, $\boldsymbol U$ is the velocity field, $\boldsymbol U_0$ is the grid velocity (i.e., the velocity which is used to maintain the shell masses, see appendix \ref{appendix:SPHERLS}) and  $S^\star = 2\pi/\left( \theta_{\rm width}\int^{\theta_0+\theta_{\rm width}}_{\theta_0}\sin\theta {\rm d}\theta\right)$ is a scaling factor used to upscale the energy from the shell to the entire star. Here $\theta_{\rm width}$ is the angular size of the model, $R_0$ is the inner radius of the envelope, $R$ is the radius of the star, $\theta_0={\pi}/{2}-\theta_{\rm width}/{2}$ is the starting horizontal ($\theta$) coordinate of the model.

The evolution of this quantity over time gives us a picture of the state of the model. First, it shows a somewhat steady rising phase, which comes from the initial small perturbation, and after that a relaxation phase in which it reaches a quasi-steady phase in which it fluctuates but has an almost constant average in time. Essentially, we can say that the first two phases are somewhat dependent on the resolution of the model, which is related to the numerical viscosity of the given model. This can be seen in Fig.~\ref{fig:TKE-evol} right panel: The larger numerical viscosity coming from the larger horizontal cell sizes of the model {\tt v036\_6\_150x20} ($N_\theta = 20$, $\Delta\theta =0.3^\circ$) delay the rising phase. Nevertheless  all models reached the quasi-steady state after $\tau_{\rm rel}$.

\subsection{Basic statistical quantities}
\label{sec:stat_quant}
There are several methods to analyze turbulent convection. In this section, we present some of the statistical methods that are broadly used in the field of turbulence.

In the literature of numerical hydrodynamical modeling of pulsation, the meaning of the terms of ``{ergodicity}'' and ``{ensemble average}'' are slightly  different  than in other branches of physics, for example in statistical physics \citep{kupka}. In this type of numerical solution, the computation time has to cover many periods or characteristic time intervals to provide good enough statistics, which requires a long computation time. Another possibility is that one runs several copies of the model, and instead of taking the average of one long sample, one takes the average of these copies. The ergodic hypothesis of numerical turbulence states that these two methods give identical results. 

Hence, to collect statistical data about the turbulent convection, we assume ergodicity; so, ensemble averages are taken as time averages of $\tau_{\rm stat} = 9$-day interval  (i.e.,  instead of averaging a static $q$ quantity of $N$ number of simulations, we take the time average of a single simulation) .  The ensemble averages  are denoted by
\begin{equation}
    \langle q \rangle = \frac{1}{N} \sum_i^N q_i \approx \frac{1}{\tau_{\rm stat}} \int_{t_0}^{t_0+\tau_{\rm stat}} q(t) {\rm d}t\text{.}
\end{equation}

In different cases, we use horizontal Reynolds and Favre averages. The former is denoted by the overbar,
\begin{equation}
    \overline{q} \equiv \frac{1}{\Delta \Omega} \int\limits^{\Delta\Omega} q(r,\theta,\phi,t) {\rm d}\Omega = \frac{\int\limits^{\theta_{\rm width}} q(r,\theta,t) \sin \theta {\rm d} \theta}{\int\limits^{\theta_{\rm width}} \sin\theta {\rm d}\theta},
\end{equation}
and the Favre average is defined by the density-weighted Reynolds average and denoted by a tilde:
\begin{equation}
    \tilde{q} = \frac{\overline{\rho q}}{\overline{\rho}}\text{.}
\end{equation}

We note that the ensemble and horizontal averaging are commutable. To mitigate the fluctuations in the statistics caused by numerical uncertainties, we use ensemble averages of the horizontally averaged quantities as well,  a procedure is widely used in the field \citep{kupka}. Using the above averaging methods, we can split quantities into mean and fluctuating parts:
\begin{equation}
    q = \overline{q} + q^\prime = \tilde{q} + q^{\prime\prime}\text{.}
\end{equation}

We note that using fluctuating quantities allows the computation of other statistical moments: for instance, the horizontal standard deviation of a quantity $q$ is defined by $\overline{\left(q^{\prime}\right)^2}$.

The Favre and Reynolds averaging and their fluctuations are related to each other as
\begin{align}
    \tilde{q} = \overline{q} + \frac{\overline{\rho^\prime q^\prime}}{\overline{\rho}} \text{;}\\
    q^{\prime\prime} = q^\prime - \frac{\overline{\rho^\prime q^\prime}}{\overline{\rho}}\text{.}
\end{align}
One can see that in the case of $\rho^\prime { /\rho} \ll 1$, the additive terms on the righthand side disappear, thus the two averaging and corresponding fluctuations are equivalent. These averages are usually used to analyze the multidimensional models in 1D and also to construct 1D models \citep{Viallet2013}.

To tackle the turbulent nature of stellar convection, we use statistical quantities usually used to describe turbulence. The structure of the velocity field can be described by the auto-correlation function.\footnote{Strictly speaking, auto-correlation is used to describe homogeneous and isotropic turbulence, while large-scale convective motions are anisotropic, homogeneity can be broken at boundary layers, and even far from the surface (see, e.g., \citealt{Hanson2024}).} The auto-correlation function at position $\boldsymbol x$ is the ensemble average of the velocity fluctuations and the fluctuation at $\boldsymbol{x}+\boldsymbol{r}_d$, where $\boldsymbol{r}_d$ is a displacement vector \citep{Pope-konyv}:
\begin{equation}
    R_{ij}(\boldsymbol{r}_d) = \langle u_i(\boldsymbol x + \boldsymbol r_d) u_j(\boldsymbol x) \rangle\text{.}
\end{equation}
Here $u_i= U_i - \langle U_i \rangle$ is the ensemble fluctuation of the velocity in $i$ direction. Since most Reynolds-averaged models used in the
pulsation-literature  \citep[see][]{Baker1987}  assume horizontal symmetry, and we  mostly consider the horizontal ($\theta_d$) displacements, $\boldsymbol{r}_d = r\theta_d \boldsymbol{e}_\theta$. On the other hand, numerical fluctuations break the strict homogeneity; thus we use the Reynolds averages as well: $\overline{R_{ij}}$. 

\begin{figure*}
    \centering
    \includegraphics[width=\textwidth]{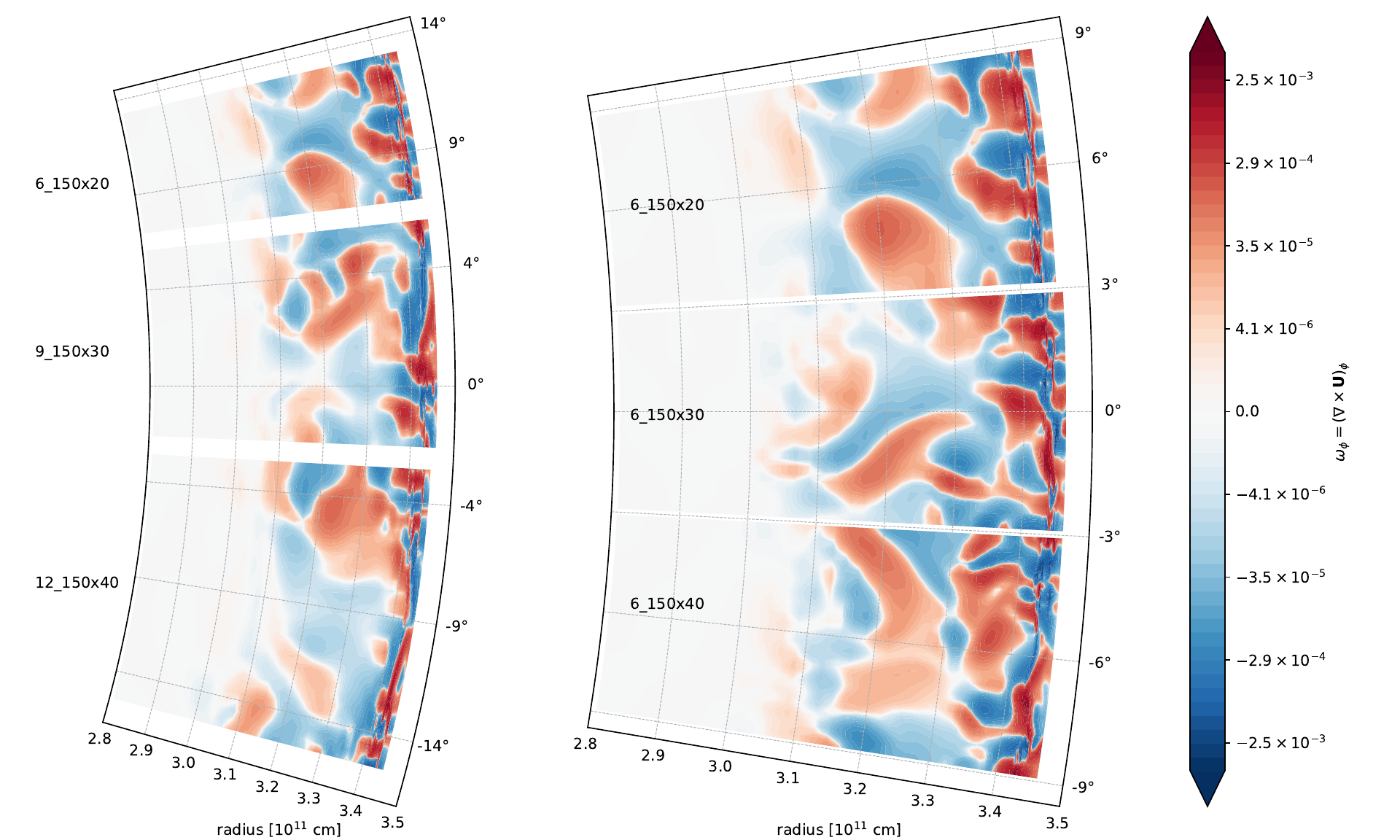}
    \caption{Snapshot of five different simulations, taken at the same simulation time. 
    The 2D simulations are in polar coordinates; the radial coordinate is in $10^{11}$ cm units; and the angular coordinates are in degrees. The color-coding denotes the vorticity of the fluid flow  The simulation domains are rotated to fit on the figure.
    Left panel: Models {\tt v036\_6\_150x20} (top), {\tt v036\_9\_150x30 } (middle), and \texttt{v036\_12\_150x40} (bottom). These models have different angle widths, but the same resolution. Right panel: \texttt{v036\_6\_150x20} (top), 
    {\tt v036\_6\_150x30} (middle), \texttt{v036\_6\_150x40} (bottom). These models have the same angle width, but   different horizontal resolutions.
    In the left panel we can see that with increasing opening angle, the number of structures is also increasing in the HI region (at around $3.4\times 10^{11}$ cm radius.). In the HeII zone (at around $3.2\times 10^{11}$ cm), the size of the  vortices  increase. This means that the initial model with of $6^\circ$ degrees used by \cite{SPHERLSII} is inadequate to properly capture the properties of the flow structure in this area.
    Meanwhile, we can see on the right-hand side that the number of structures in the HI region is similar with higher resolutions (although  more details appear).}
    \label{fig:vorticity}
\end{figure*}

We show vorticity ($\omega = {\rm rot} (\boldsymbol{U}-\boldsymbol{U}_0)$) maps of different models in Fig.~\ref{fig:vorticity}. All models are different realizations of v036. The left panel shows realizations with the same resolution and different widths, while the right panel shows different resolutions with the same angular width. We note that we 
included an animated version of this figure as supplementary material, in which one can follow the change of the vorticity field in $\tau_{\rm stat}$ time. In this video, one can see that the different blobs move upward and downward, and there are larger ones that also rotate. To quantify the characteristic length scales of these eddies, we use the integral length scale:
\begin{equation}
     L_{ij} = \int\limits_0^{r_{\max}} R_{ij} (r_d)/R_{ij}(0) {\rm d}r_d\text{.}
\end{equation}

 In engineering cartesian scenarios, the upper boundary $r_{\max}$ is set to infinity \citep{Pope-konyv}. This cannot be maintained in stellar astrophysical scenarios, since in the radial direction it makes no sense to continue the integration into the interstellar medium, while in the horizontal domain the star has a finite extent of $2\pi R$, which is a periodic boundary resulting in a self interaction. In our case the horizontal extent is much smaller being $\theta_{\rm width} R$. The periodic boundary also modifies the shape of the autocorrelation function, thus we select $r_{\max}$ based on the following conditions \citep{ONeill2004}:
\begin{itemize}
    \item we integrate until the auto-correlation function reaches zero,
    \item or until it reaches a minimum value.
\end{itemize}
These statistics give us a quantity that can be dependent on the horizontal width and resolution. If the angular size of the model is too small, it prevents the creation of larger eddies.  The nonlinear interaction between convection and pulsation causes cycle-to-cycle variations in the pulsation (alongside with granulation noise), which can be severely modified by unphsysical self-interaction caused by the too small horizontal size. Having the current limitations regarding $\theta_\text{width}$, we use $L_{ij}$ to minimize this effect, as this length scale does not increase further after reaching a minimum optimal angular size \citep{ONeill2004}.

In addition to the length scale of the eddies, the overall horizontal fluctuation of the velocity field (or root-mean-square difference velocity) is an important quantity, as it is used directly in the 1D pulsation codes, so it can be compared to 1D results. Hence, it is crucial to map this dependence on the horizontal model size, $\theta_{\rm width}$ and horizontal resolution, $\Delta\theta$. We define these velocities by:
\begin{align}
   & u_{r, \rm rms} \equiv \sqrt{\left\langle\overline{\left(U_r^\prime\right)^2} \right\rangle} = \sqrt{2e_{22}}\text{,}\\
    &    u_{\theta, \rm rms} \equiv \sqrt{\left\langle\overline{\left(U_\theta^\prime\right)^2} \right\rangle} =\sqrt{2e_{11}}\text{.}
\end{align}

We note that by definition, these are equivalent to the square root of  twice  the horizontal and radial specific turbulent kinetic energy densities ($e_{11}$ and $e_{22}$ , respectively).\footnote{We use $e_{11}$ for the horizontal kinetic energy. We chose this notation because we only consider horizontal displacements in $r_d$; hence, the longitudinal integral length scale is based on the horizontal velocities. This latter is noted $L_{11}$ by convention; therefore, we chose $i=1$ to be the $\theta$ direction and $i=2$ the radial direction} The total specific kinetic energy is defined by
\begin{equation}
    e_{t,0} = e_{11} + e_{22} = \frac{1}{2}\left\langle \overline{(U_r^\prime)^2 + (U_\theta^\prime)^2} \right\rangle\text{.}
\end{equation}

Since the Rayleigh-Bénard cells of the convection are genuinely anisotropic, and these anisotropies have their effects in the pressure terms of the turbulence \citep{Pope-konyv}, the large scale anisotropy of the flow is also interesting. This is defined by the ratio of the rms velocities:
\begin{equation}
    \sigma_{r\theta} \equiv \frac{u_{r,\rm rms}}{u_{\theta,\rm rms}}\text{.}
\end{equation}

 Convection in the stellar matter is always turbulent, as the viscous forces are negligibly small \citep{kupka}. Meanwhile, the radiative diffusion is large and the Prandtl numbers are in the $10^{-7}$-$10^{-10}$ range. Turbulent convection with these low Prandtl numbers follows Kolmogorov-scaling \citep{Verma2011}, meaning that kinetic energy of the large scale convective motions are transferred through a cascade of eddies to scales where viscosity is relevant \citep{Pope-konyv}. In this case, the kinetic energy in the cascade follows a power law of $-5/3$. Thus, the spatial spectra of the turbulent kinetic energy can be divided into three spatial ranges: the energy-containing range, which contains the bulk \citep[>80\%][]{Pope-konyv} of the kinetic energy; the inertial subrange follows the Kolmogorov-law and the dissipation range, where the kinetic energy decreases exponentially, and is transformed into heat. In LES simulations, the main focus is to resolve the energy-containing range, which is done by a filtering function; however, this filter affects damping energies. To minimize the damping effect, one has to maintain a filter resolution in which $\Delta_{\rm filt} \le L_{11}$. In our case $\Delta_{\rm filter} = l = \sqrt{r\Delta r\Delta\theta}$ \citep{Magnient2007}.
 
 The other source of damping is numerical diffusion or numerical viscosity, which is caused by the truncation errors of the numerical scheme of the code. This actually can be more severe than the damping effect of the filter (see appendix \ref{appendix:numerics} and \ref{appendix:tests}). Therefore, it is worth to study the energy cascade of these models and compare it to the literature.

To quantitatively describe the energy cascade, we define the  energy spectrum tensor  by the Fourier transform of the auto-correlation function \citep{Pope-konyv},
\begin{equation}
\label{eq:energy-spec}
    \Phi_{ij}(\kappa) = \frac{1}{\mathcal L} \int\limits_0^\mathcal{L} R_{ij}(r_d,t) e^{-i\kappa r}{\rm d}r_d\text{,}
\end{equation}
where $\mathcal L$ is the horizontal width of the model $\mathcal{L} = r  \theta_{\rm width}$
With having $N$ horizontal cells, the grid spacing horizontally is $\theta_{\rm width}/N_\theta=\Delta \theta$, which translates to  a cut-off scale 
 of $r\Delta \theta \equiv \Delta_{\rm c}$. In practice, the largest horizontal frequency (Nyquist-frequency) is at wave number $\kappa_c = \pi/\Delta_{\rm c}$, and one can obtain $N/2$ number of spectrum points. Following this concept, the spatial spectrum of the turbulent kinetic energy is
\begin{equation}
    e_t(\kappa) = \frac{1}{2} (\Phi_{11}(\kappa) + \Phi_{22}(\kappa))\text{.}
\end{equation}

We note that by using horizontal means, all previously described quantities will be dependent on the radius coordinate inside the star.

\subsection{Dimensionless quantities}
\label{sec:dimless}
Although we do not have direct measurements of RR Lyrae subsurface convection, one can derive dimensionless numbers from material physics to characterize flow in these layers \citep{kupka}.\footnote{In the case of the Sun, the observations give a map of the top layer of convection cells with $\sim10$~km resolution, and the numerical  resolution can go down a bit below this.}
The other and commonly used approach is comparing modeling results to observations. However, we need to follow many pulsation periods in nonlinear radial pulsation modeling, which has an unfeasible numerical cost in the case of two or three dimensional computations. One would need multiple observed stars, and for each, a dozen models followed through at least 400--500 cycles, which calculation for a single model can take months ( for the computational cost in SPHERLS, see Section~\ref{sec:numcost} and also the original SPHERLS papers of \citealt{SPHERLS1,SPHERLSII,SPHERLSIII,SPHERLS4}). Hence, we use dimensionless numbers that characterize the flow and underlying numerics.

The dimensionless numbers characterizing the flow are the ratio of effective kinematic to radiative energy diffusion (effective Prandtl number, $Pr_{\rm eff}$), the ratio of inertial to viscous forces (Reynolds number, $Re$), the ratio of total energy flux to radiative energy transport (Nusselt number, $Nu$) and the ratio of convective velocities to radiative diffusion (Péclet number, $Pe$). We also define the horizontal Reynolds number ($Re_\text{hor}$) in which we only consider horizontal velocities and horizontal viscosity. 

Their definitions:

\begin{displaymath}
       Pr_{\rm eff} = \frac{\nu_{\rm num} + \nu_{\rm SGS}}{\chi_\text{tot}},\quad Re = \frac{u_{\rm rms}2L_{11}}{\nu_{\rm num}+\nu_\text{SGS}},\quad 
\end{displaymath}
\begin{displaymath}
   {  Re_{\rm hor} = \frac{u_{\theta,{\rm rms}}2L_{11}}{\nu_{{\rm num},\theta}+\nu_\text{SGS}},}\quad Nu=\frac{F_r+F_c}{F_r},\quad  Pe=\frac{u_{\rm rms}2L_{11}}{\chi_\text{tot}},
\end{displaymath}
where $\nu_{\rm num}$ is the kinetic diffusivity of the numerics,  $\chi_\text{tot}=\chi + \chi_{\rm SGS}+\chi_\text{E}^{\rm num}+\chi_\text{T}^\text{num}$, with $\chi=16\sigma T^3/(3\kappa c_p\rho^2)$ which is the thermal diffusivity. The quantities $\chi_\text{E}^{\rm num}$ and $\chi_\text{T}^\text{num}$ are the internal energy and heat diffusivity of the numerics, while $\chi_{\rm SGS}=\nu_{\rm SGS}/(Pr_t)$ which is the subgrid scale model diffusivity with $\nu_{SGS}$ as the SGS viscosity and $Pr_t$ is a constant (see Appendix~\ref{appendix:SPHERLS}).  The characteristic length scale of the flow is chosen to be the size of the eddies, $L_{11}$. The numerical diffusivities are discussed in detail in Appendix~\ref{appendix:numerics}.

The last undefined term in the above definitions is $F_c$, the convective flux. This can be described as the covariance of the velocity and enthalpy ($h$) fields:
\begin{equation}
\label{eq:F_c}
    F_c = \langle \overline{\rho} \widetilde{h^{\prime\prime} U_r^{\prime\prime}}  \rangle\text{.}
\end{equation}

To measure the effects of the numerical methods and the resolutions, we define the following quantities: the ratio of resolved to SGS kinetic energies ($q_{\rm SGS}$), the ratio of the numerical  to SGS viscosities
($q_{\rm num}$) and the ratio of the numerical energy diffusivity to SGS energy  diffusivity ($q_{\rm diff}$):
\begin{displaymath}
    \quad{   q_{\rm SGS} = \frac{e_{\rm SGS}}{e_{t,0}}}, \quad q_{\rm num} = \frac{\nu_{num}}{\nu_{\rm SGS}},\quad q_{\rm diff} = \frac{ \chi_\text{E}^{\rm num}+\chi_\text{T}^\text{num}}{\chi_{\rm SGS}} \text{,}
\end{displaymath}
where $\chi_{\rm num}$ is the numerical energy diffusivity (see Appendix \ref{appendix:numerics}).

A  significant problem of stellar convection modeling is that we cannot directly simulate $Pr\sim 10^{-7}$ flows, as the numerical dissipation would act as an additional viscosity and increase the effective Prandtl number. In SPHERLS, the numerical viscosity strongly depends on the model resolution and the current flow velocities  (see Appendix~\ref{appendix:numerics}); therefore, it depends on location, and we cannot determine it before calculations. To circumvent this problem  and describe expected flow characteristics before simulation, we define a characteristic Prandtl and Reynolds number that is used to describe expected Prandtl and Reynolds number differences via different zoning and resolution. For this, we use the minimum effective numerical viscosity needed for stable solution of hyperbolic partial differential equations \citep{Strikwerda2004}, in the form of
\begin{equation}
    \nu_{\rm eff} = \frac{\Delta U_{\rm ch} l_{\rm ch}}{2}\text{,}
\end{equation}
where $\Delta U_{\rm ch}$ is the characteristic  velocity differences on the grid scale ($1.5$ km/s expected at this resolution with $U_\text{ch}\approx10$ km/s characteristic velocities for RR Lyrae stars based on 1D calculations), and $ l_{\rm ch}$ is the characteristic cell size of the model taken in the HI ionization zone.   Since the radial grid is finer in the HI zone, we set $l_{\rm ch}=R\Delta\theta$. The characteristic length scales of the eddies, $L_{\rm ch}$ is chosen to be $1.5H_p$, based on the 1D mixing length theory, and usual parametrization of 1D codes \citep{Paxton2019} The radiative thermal  diffusivity $\chi_{\rm ch}$,  and pressure scale height $H_p$ is calculated from 1D models at 10 000 K. So, our characteristic Prandtl number is defined as
\begin{equation}
\label{eq:Pr_ch}
     Pr^{\rm ch} = \frac{\nu_{\rm eff}}{\chi_\text{ch}} =\frac{\Delta U_{\rm ch}R \Delta \theta}{2\chi_{\rm ch}}, 
\end{equation}
 and the Reynolds number as
\begin{equation}
    \label{eq:Re_ch}
    Re_\text{ch} = \frac{U_\text{ch}L_c}{\nu_{\rm eff}}=\frac{3U_\text{ch}H_p}{\Delta U_{\rm ch}R \Delta\theta}\text{.}
\end{equation}

 We  use these quantities to get a predominant view about the expected behavior of the simulations, as in this form they only depend on $\Delta r$ and $\Delta \theta$ which are defined by our choice of cell numbers ($N_r$ and $N_\theta$ respectively) and the horizontal size of the domain ($\theta_\text{width}$) in the latter.

The final goal of our study is the comparison of the 1D and 2D pulsation models; hence, it is crucial to determine how the 1D properties of the convective regions depend on the domain size and resolution.
We define five length scale parameters characterizing the size of the different structural components of the convective zone in RR Lyraes; see Table~\ref{tab:scales} and Table~\ref{tab:scales2}. 

These length scale parameters are given in pressure scale height units ($H_p=pr^2/(\rho G M_r)$, where $p$ is the pressure, $M_r$ is the mass inside the sphere with radius $r$ and $G$ is the gravitational constant). We use the $H_p$ at the bottom of the HeII partial ionization zone.

\begin{table}[ht!]
    \centering
    \caption{Terminology of the convective structure components used in this paper.}
    \label{tab:scales}
    \begin{tabular}{ccc}
            \hline
        \hline
        Name & Condition & Buoyancy role \\
        \hline
        Convective zone &  $F_c>0$ ; $Y>0$ & Acceleration \\
        Counter-gradient & $F_c>0$ ; $Y<0$ & Ineffective deceleration \\
        Overshooting  &  $F_c<0$ ; $Y<0$ & Effective deceleration \\
        {\small Non-zero turbulence} & $e_{t,0} > 10^6$ erg & N/A\\
    \end{tabular}
    \tablefoot{The third column describes what role the buoyancy plays in the given region. The nonzero turbulence region is not connected directly to a given role.}
    
\end{table}

\begin{table}[ht!]
    \centering
    \caption{Length scales corresponding to the convective structure components used in this paper.}
    \label{tab:scales2}
    \begin{tabular}{cl}
        \hline
        \hline
        Notation & \multicolumn{1}{c}{Meaning} \\
        \hline
        $\Lambda_{\rm conv}$ & The entire convective region\\
        $\Lambda_{\rm HI}$ & Conv. region in HI part. ion. zone \\
        
        $\Lambda_{\rm HeII}$ & Convective region in HeII part. ion. zone \\
        $\Lambda_{OV}$ & Overshooting zone\\
        $\Lambda_t$ &  Region of non-zero turbulence \\

    \end{tabular}
    
\end{table}

The length scales are defined by the relation of the convective flux $F_c$ and the nondimensional super-adiabatic gradient $Y=-(H_p/c_p)(\partial s/\partial r)$ similarly to \citet{Kapyla2019}, where $c_p$ is the specific heat at constant pressure and $s$ is the entropy. The length scales $\Lambda_{\rm HI}$, and $\Lambda_{\rm HeII}$ are defined by the buoyancy-driven convectively unstable region, that is, where $F_c>0$ and $Y>0$. We define the so-called Deardorff's or counter-gradient\footnote{This is a variant of overshooting 
generated by the nonlocal nature of the turbulent flow
where flux transported into stable layers (see, e.g., \citet{Kupka2020} section 5). The detailed features of these phenomena and the layer will be discussed in the following papers.} regions by the condition of $F_c > 0$ and $Y < 0$. The overshooting zone, $\Lambda_{\rm OV}$, is described by the negative convective flux, that is $F_c < 0 $ and $Y<0$. We note that since $F_c\rightarrow0$ asymptotically, we have an extra condition as $|F_c|/|\max F_c|> 0.001$. The length scale of the size of the convective region is $\Lambda_{\rm conv}=L_c/H_p$, where $L_c$ is the size of the convective region, which is defined from the bottom of the HeII ionization zone ($R_{\rm bottom}$) to the top of the HI ionization zone ($R_{\rm top}$): $L_c = R_{\rm top}-R_{\rm bottom}$ and $u_{\rm rms}=\sqrt{u_{\theta,\rm rms}^2+u_{r,\rm rms}^2}$. The effectively turbulent region, $\Lambda_t$ , is defined by the condition $e_{t,0} \ge 10^6$ erg. This is an important length scale to compare with the 1D calculations, as the 1D turbulent convection models of radial stellar pulsations are explicitly solving equations for $e_{t,0}$, and many other terms of these models depend on this quantity\footnote{ Including the source term of the turbulent convection \citep{Kuhfuss1986}. To have actually $e_{t,0}$ in the models, a so-called minimum value is added numerically to the system \citep{Yecko1998}. There are fundamental differences between 1D $e_{t,0}$ profiles of the different codes \citep[see the discussion of][]{lengyel}, hence we chose the minimum value of accepted turbulent energy level of $10^6$ erg, which is 2-4 magnitudes higher than the minimum value used in 1D hydrocodes.} \citep{bpf-beat2002,Paxton2019,KovacsGB2023}.

Due to the choice of the $H_p$ parameter and the fact that the two buoyancy-driven convective regions are separated, the length scales have this expected relation:
\begin{displaymath}
    \Lambda_t > \Lambda_{\rm conv} > \Lambda_{\rm OV} \gtrsim \Lambda_{\rm HeII} > \Lambda_{HI}\text{.}
\end{displaymath}

Alongside with the length scales, we also define the mean convective velocity, $u_{\rm rms}^{\rm tot}$, and the convective turnover timescale $\tau_{\rm conv}$.  These values give us an overall estimate about the timescale of convective processes, and the total kinetic energy stored in the convective region, and were used for a red giant model by \citet{Viallet2013}.
The mean convective velocity, $u_{\rm rms}^{\rm tot}$, is defined similarly to \cite{Arnett2009},
\begin{equation}
    \frac{1}{2} M_{\rm conv} (u_{\rm rms}^{\rm tot})^2 = E_{\rm t,conv}\text{,}
\end{equation}
where $E_{\rm t,conv}$ is the total turbulent kinetic energy in the convective shells, and $M_{\rm conv}$ is the total mass of the convective shells: 

\begin{align}
    &E_{\rm t,conv} = 2\pi\int\limits^{R_{\rm top}}_{R_{\rm bottom}} \int\limits_{\theta_0}^{\theta_0+\theta_{\rm width}} \rho (r,\theta)\, |\boldsymbol U(r,\theta) -\boldsymbol U_0(r)|^2 r^2 \sin\theta {\rm d}\theta {\rm d}r   \text{;}\\
    &M_{\rm conv} = 2\pi\int\limits^{R_{\rm top}}_{R_{\rm bottom}} \int\limits_{\theta_0}^{\theta_0+\theta_{\rm width}} \rho (r,\theta) r^2\sin\theta {\rm d}\theta {\rm d}r\text{.}
\end{align}

This is used to define the convective turnover timescale: $\tau_{\rm conv} = L_c/u_{\rm rms}^{\rm tot}$.

In summary, we use different statistical and horizontal averaged quantities that describe the size of the turbulent eddies, the behavior of the velocity field, the turbulent cascade, general flow properties, and the size and behavior of the convective region. We are interested in determining whether these quantities are affected by the sizes and the resolutions of the 2D models, and if they are, then which model parameters give consistent results, and what is the best compromise between resolution and numerical performance. In the next section, we first show the fundamental computational limitations of the calculations, and then the different results of the aforementioned quantities. 

\section{Results}
\label{sec:results}
\subsection{Calculation times}
\label{sec:numcost}
In this section, we present some benchmark information that are relevant to our study. The main reason behind resolution tests is the calculation times. We aim at finding a model resolution that:
\begin{itemize}
    \item[a)] can resolve horizontal scales that are relevant in radial pulsation physics
    \item[b)] can ensure reaching full amplitude pulsations within reasonable computation times.
\end{itemize}
Therefore, here we study the evolution of time steps in the different scales and the scaling of computation times on different numbers of cores and different resolutions on two different machines. 

\begin{figure}[htb!]
    \centering
    \includegraphics[width=\columnwidth]{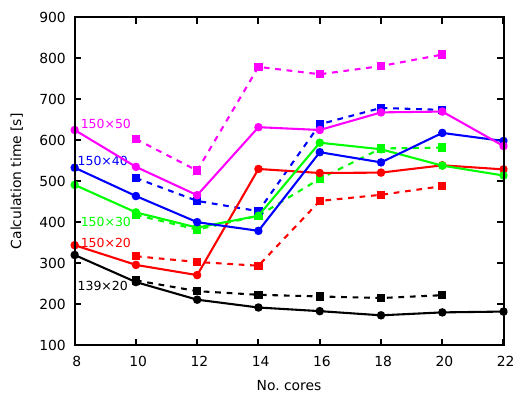}
    \caption{Calculation times for $10\,000$ time steps, on HUN-REN Cloud (dashed line, filled squares) and the HUN-REN Malor computer (solid line, filled circles). The API overhead becomes a problem after 12 cores, while in the low vertical resolution case, {this happens only for} 18 cores. }
    \label{fig:comput-times}
\end{figure}

In Fig.~\ref{fig:comput-times}. we show the calculation time on different number of cores in two different environments. The filled squares represent calculations on the HUN-REN Cloud \citep{H_der_2022}, with AMD EPYC-Rome CPU, the filled circles are points calculated on the HUN-REN CSFK Malor computer with a CPU of Intel Xeon Gold 5220R CPU @ 2.20GHz.  One can see that in most cases, after reaching twelve cores, the computational time increases, which is a phenomenon called API overhead. This problem is caused here by the OpenMPI environment, in which parallelization is solved by having the parallel threads in different processes. These processes have their own memory and do not see into each other's, hence, a massage interface is used to communicate between the different parallel processes. This allows efficient usage of large CPU clusters. However, in this case, due to the continuous synchronizing between cores in each time step, after (in this case) twelve parallel processes, the communication between cores takes more time than the calculation.

Since SPHERLS uses an explicit method to update velocities \citep{SPHERLS1}, the time steps must be kept below the Courant-criterion. This causes two problems: first, as we reach the surface, the pressure scale height drops, decreasing the sound speed and thus the time step; secondly, the better horizontal resolution also decreases the time step for a given velocity. These effects can decrease time steps to the level of $\Delta t \sim 10^{-3}$ seconds, which means that in the case of 10 000 time steps, one propagates the model by 10 seconds only, but to do this the calculation time amounts to 8 minutes.

\subsection{Structure of the convective zone}
\label{sec:conv_struct}

\begin{figure*}[htbp!]
    \includegraphics[width=\textwidth]{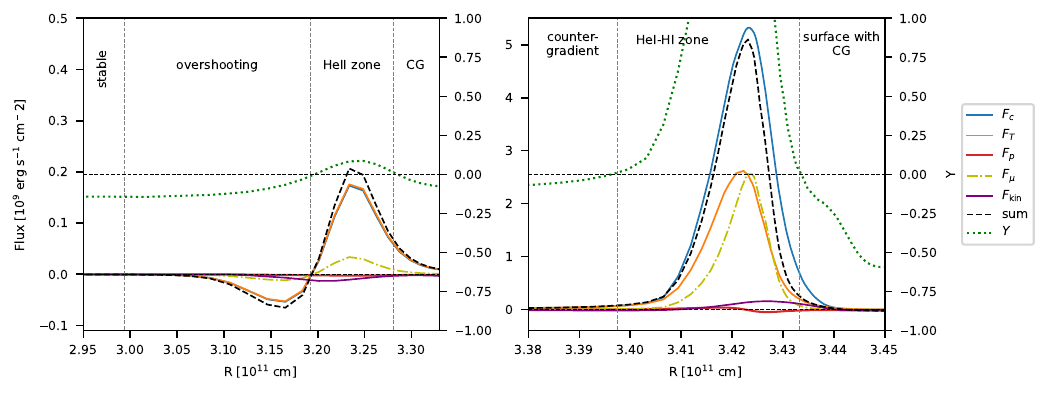}
    \caption{Structure of the convective zone and the flux components. The left-hand side is the vicinity of the HeII zone, and the right-hand side is the vicinity of the HI zone. The left-hand side vertical axis is the flux, and the horizontal axis is the radius. The green dotted curve with the right-hand axis denotes the dimensionless entropy gradient. Comparing the sign of this and the convective flux (blue curve) gives the borders of the different regions, which are separated by the dashed gray vertical lines. The condition of the regions is given in Section~\ref{sec:methods}, and also in Table \ref{tab:scales}. The other flux components are the temperature flux (orange curve), the pressure flux (red curve), and the ionization energy flux, $F_\mu$ (yellow dash-dotted curve). The sum of these three components gives the black dashed curve. The kinetic energy flux is shown as a purple line.}
    \label{fig:conv_struct}
\end{figure*}

The convective zone in the observed quasi-static RR Lyrae stellar models can be divided into four main regions based on the conditions listed in Table \ref{tab:scales}. This structural division is presented together with the components of the convective flux in Fig.~\ref{fig:conv_struct}. The zones can be followed through the relation of the super-adiabatic (or dimensionless entropy) gradient, $Y$ (presented by a green dotted line, values corresponding to the right vertical axis), and the convective flux, $F_c$ (presented by the blue curve, values belonging to the left vertical axis). The identified regions are the convectively unstable zone at the second helium ionization zone (we denote it with HeII), the unstable zone in the overlapping hydrogen and first helium partial ionization zones (HI), a counter-gradient layer (CG) which connects the two unstable zones, and a large overshooting region (OV) penetrating into the lower layers of the star.  Above HI, one can also see a thin counter-gradient and overshooting region, which are very thin compared to the regions below.

In the animated version of Fig.~\ref{fig:vorticity}, one can also see plumes in the CG region, originating from the HI region and penetrating into the HeII region. The HeII region has larger rotating structures, and blobs from these can break off, escaping in either direction. Those in the CG usually connect into the HI region, while in the OV they are dissolved. 

\begin{figure}[htb!]
    \centering
    \includegraphics[width=\columnwidth]{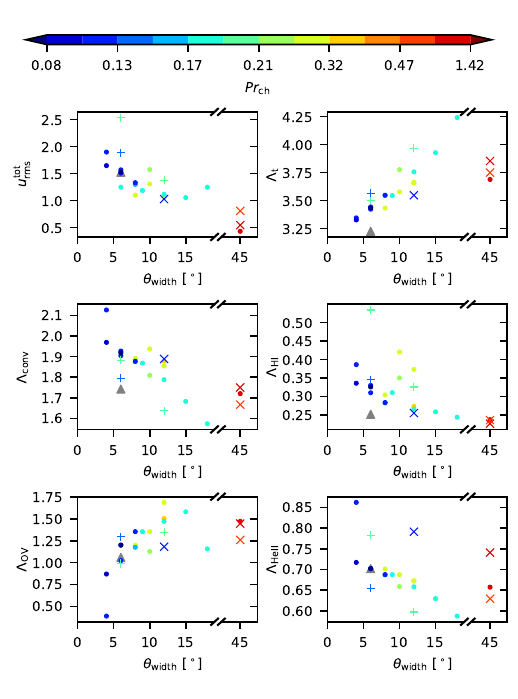}
    \caption{Different properties of convection vs.  the horizontal size of the model. These data are also presented in Table~\ref{tab:convective_properties}. The color-coding denotes the resolution through the characteristic Prandtl number. The standard vertical resolution v036 models are denoted by circles, and  the reduced vertical resolution ($N_r=122$) models by  crosses. Pluses denote the v046 runs, and the {\tt v036\_6\_200x20} model is represented by the gray triangle. Top left panel: Average convective velocity; top right panel: Size of the turbulent region ($\Lambda_t$) in pressure scale height ($H_p$) units. Middle left: Size of the total convective region in $H_p$ units. Middle right: Size of the hydrogen ionization zone in $H_p$ units. Bottom left: Size of the overshooting region in $H_p$ units. Bottom right: Size of the second helium ionization zone in $H_p$ units. }
    \label{fig:conv-prop}
\end{figure}

The sizes of the different layers for every model are given in Table~\ref{tab:convective_properties}. We plot these and $u_{\rm rms}^{\rm tot}$ against the $\theta_{\rm width}$ in Fig.~\ref{fig:conv-prop}. The color code denotes the characteristic Prandtl numbers of the different models. Models of v046 are denoted by plus signs, and models with radial cell numbers of 122 are denoted by crosses; \verb|v036_6_200x20| is presented by the gray triangle. For $u_{\rm rms}^{\rm tot}$, $\Lambda_{\rm conv}, \Lambda_{\rm HI}$ and $\Lambda_{\rm HeII}$, there is a decreasing trend with the horizontal width, while $\Lambda_{\rm OV}$ and $\Lambda_{t}$ increases when we increase the $\theta_{\rm width}$. While in the $\theta_{\rm width} < 10^\circ$ region, the resolution has larger effects than in the larger $\theta_{\rm width}$ regime.

The consistency and amplitude of differences are also different for the different $\Lambda$s. The largest change happens with $\Lambda_{OV}$, changing by a factor of 3. This means that $\theta_{\rm width}>6^\circ$ is a crucial condition for having meaningful simulations in the HeII region. The models with low resolution ($Pr_{\rm ch} > 0.3$) show very low velocities, but otherwise, they fit in the trend in the case of the other parameters. v046 models show the same trends as v036 models, while the vertical resolution has a small effect.

 Interestingly, we did not find a counter-gradient region between the overshooting and HeII zones, not even in the higher resolution models. In low-resolution models, the CG zone can be "squeezed out", meanwhile convection is already insufficient in the HeII region, causing an insufficient transport of flux downward. On the bottom-line, this feature needs further investigation in quasi-static RR Lyrae models in the future.

\begin{table*}[htbp!]
    \centering
    \caption{Horizontal averaged convective properties of the models.}
    \label{tab:convective_properties}
    \begin{tabular}{lcccccccc}
    \hline
    \hline
    Name & $u_{\rm rms}^{\rm tot}$ [km/s]& $\tau_{\rm conv}$ [day] & $H_P$ [cm]& $\Lambda_t$ & $\Lambda_{\rm conv}$ & $\Lambda_{\rm HI}$ & $\Lambda_{\rm HeII}$ &$\Lambda_{\rm OV}$ \\
    \hline
\verb|v036_4_150x20  | & $1.90$ & $3.21$ & $1.24\times 10^{10}$ & $3.35$ & $2.13$ & $0.39$ & $0.86$ & $0.39$ \\
\verb|v036_4_150x30  | & $1.65$ & $3.42$ & $1.24\times 10^{10}$ & $3.33$ & $1.97$ & $0.34$ & $0.72$ & $0.87$ \\
\hline
\verb|v036_6_150x20  | & $1.25$ & $4.50$ & $1.26\times 10^{10}$ & $3.44$ & $1.92$ & $0.33$ & $0.70$ & $1.02$ \\
\verb|v036_6_150x30  | & $1.51$ & $3.74$ & $1.26\times 10^{10}$ & $3.44$ & $1.93$ & $0.33$ & $0.70$ & $1.02$ \\
\verb|v036_6_150x40  | & $1.57$ & $3.55$ & $1.26\times 10^{10}$ & $3.42$ & $1.90$ & $0.31$ & $0.70$ & $1.20$ \\
\verb|v036_6_150x50  | & $1.53$ & $3.67$ & $1.26\times 10^{10}$ & $3.44$ & $1.92$ & $0.33$ & $0.70$ & $1.20$ \\
\verb|v036_6_200x20  | & $1.52$ & $3.42$ & $1.29\times 10^{10}$ & $3.23$ & $1.74$ & $0.25$ & $0.70$ & $1.06$ \\
\hline
\verb|v036_8_150x20  | & $1.10$ & $5.02$ & $1.26\times 10^{10}$ & $3.43$ & $1.89$ & $0.30$ & $0.70$ & $1.20$ \\
\verb|v036_8_150x30  | & $1.30$ & $4.31$ & $1.29\times 10^{10}$ & $3.55$ & $1.88$ & $0.28$ & $0.69$ & $1.17$ \\
\verb|v036_8_150x40  | & $1.33$ & $4.21$ & $1.29\times 10^{10}$ & $3.55$ & $1.88$ & $0.28$ & $0.69$ & $1.35$ \\
\hline
\verb|v036_9_150x30  | & $1.19$ & $4.70$ & $1.29\times 10^{10}$ & $3.54$ & $1.87$ & $0.31$ & $0.69$ & $1.35$ \\
\hline
\verb|v036_10_150x20 | & $1.31$ & $4.43$ & $1.29\times 10^{10}$ & $3.58$ & $1.94$ & $0.42$ & $0.69$ & $1.35$ \\
\verb|v036_10_150x30 | & $1.57$ & $3.58$ & $1.35\times 10^{10}$ & $3.78$ & $1.81$ & $0.35$ & $0.66$ & $1.13$ \\
\hline
\verb|v036_12_122x60 | & $1.03$ & $5.53$ & $1.30\times 10^{10}$ & $3.55$ & $1.89$ & $0.25$ & $0.79$ & $1.18$ \\
\verb|v036_12_150x20 | & $1.09$ & $5.26$ & $1.32\times 10^{10}$ & $3.66$ & $1.88$ & $0.27$ & $0.67$ & $1.50$ \\
\verb|v036_12_150x30 | & $1.05$ & $5.38$ & $1.32\times 10^{10}$ & $3.66$ & $1.86$ & $0.37$ & $0.67$ & $1.69$ \\
\verb|v036_12_150x40 | & $1.12$ & $5.00$ & $1.35\times 10^{10}$ & $3.76$ & $1.79$ & $0.27$ & $0.66$ & $1.47$ \\
\hline
\verb|v036_15_150x50|\tablefootmark{$\star$} & $1.05$ & $5.20$ & $1.41\times 10^{10}$ & $3.93$ & $1.68$ & $0.26$ & $0.63$ & $1.58$ \\
\hline
\verb|v036_18_150x60|\tablefootmark{$\star$} & $1.25$ & $4.41$ & $1.51\times 10^{10}$ & $4.24$ & $1.57$ & $0.24$ & $0.59$ & $1.16$ \\
\hline
\verb|v036_45_122x40 | & $0.54$ & $10.31$ & $1.39\times 10^{10}$ & $3.85$ & $1.75$ & $0.23$ & $0.74$ & $1.45$ \\
\verb|v036_45_122x60 | & $0.81$ & $6.47$ & $1.36\times 10^{10}$ & $3.75$ & $1.67$ & $0.24$ & $0.63$ & $1.26$ \\
\verb|v036_45_150x20 | & $0.43$ & $12.44$ & $1.35\times 10^{10}$ & $3.69$ & $1.72$ & $0.24$ & $0.66$ & $1.47$ \\
\hline
\verb|v046_6_152x20  | & $2.54$ & $2.49$ & $1.45\times 10^{10}$ & $3.50$ & $1.88$ & $0.53$ & $0.78$ & $0.99$ \\
\verb|v046_6_152x30  | & $1.88$ & $3.20$ & $1.45\times 10^{10}$ & $3.57$ & $1.79$ & $0.35$ & $0.65$ & $1.29$ \\
\verb|v046_12_152x40 | & $1.38$ & $4.34$ & $1.58\times 10^{10}$ & $3.97$ & $1.64$ & $0.33$ & $0.60$ & $1.35$ \\
    \end{tabular}
\tablefoot{The panels are separated according to the angular size and physical parameters of the models. The columns are (from left to right) the mean convective velocity ($u_{\rm rms}^{\rm tot}$), convective turnover timescale ($\tau_{\rm conv}$), pressure scale height at the bottom of the convective zone ($H_p$), turbulent length scale ($\Lambda_t$), full length of the entire convective region ($\Lambda_{\rm conv}$), length scale of the hydrogen ionization zone ($\Lambda_{\rm HI}$), length of the second helium ionization zone ($\Lambda_{\rm HeII}$), size of the overshooting region ($\Lambda_{\rm OV}$).\\
\tablefoottext{$\star$}{This model did not reach $\tau_{\rm relax}$ simulation time, due to the large numerical cost; therefore, it is not reliable. It is only included here to show the trends. Statistical evaluation is performed  over an interval of 12--18 days.}}
\end{table*}

\subsection{Components of the convective flux}
\label{sec:components}
As our final goal is the comparison with 1D models, we study the different decompositions of the convective flux, that are used to derive 1D models. The energy transported by the convective motions is typically the fluctuating enthalpy flux $F_c$ \citep[see eq. \ref{eq:F_c} and also][]{kupka}. Regularly, the two main methods using thermodynamic equities to split this term into its components  are \citep{Viallet2013}
\begin{equation}
\label{eq:flux_split_entropy}
    h^{\prime} = \overline{T} s^{\prime} + \overline{\left(\frac{1}{\rho}\right)} p^{\prime} + \overline{\left.\frac{\partial h}{\partial \mu}\right|_{s,p}} \mu^{\prime}\text{,}
\end{equation}
\begin{equation}
\label{eq:flux_split_temperature}
    h^{\prime} = \overline{c_p} T^\prime + \overline{\left(\frac{1-\delta}{\rho}\right)}p^\prime + \overline{\left.\frac{\partial h}{\partial \mu}\right|_{T,p}} \mu^{\prime}\text{,}
\end{equation}
where $\delta = (\rho/T)(\partial\ln \rho/\partial\ln T)_p$ is the isobaric compressibility, and $\mu$ is the mean molecular weight. In most cases, the pressure terms are neglected based on filtering out acoustic waves \citep[this is called low Mach number approximation][]{Gough1977,Stellingwerf1982a,Kuhfuss1986}. Eq.~\ref{eq:flux_split_entropy} is usually used in the descriptions of the current state-of-the-art pulsation codes \citep{Bono1994,Yecko1998,Paxton2019}. \citet{Viallet2013} showed that in the case of eq. \ref{eq:flux_split_entropy} the pressure term is, in fact, significant in the stellar envelope; meanwhile, in eq.~\ref{eq:flux_split_temperature}, the compressibility mostly compensates the pressure term ($\delta \approx 1$). This latter equation is used many times for analysis \citep{SPHERLSII}, and for modeling, as well \citep{Canuto1998}.

The last term of Eqs.~\ref{eq:flux_split_entropy} and \ref{eq:flux_split_temperature} are also neglected in the stellar envelopes, usually because the mean molecular weight changes only because of the ionization, which can be incorporated into $c_p$ and $\delta$ or $s$. On the other hand, in the case of the sharp stratification of the ionized zones in  RR Lyrae stars, the ionized matter that moves upward by the convective motions carries a significant amount of latent heat: their ionization energy. We found this in our study to be a significant term, giving up to 50\% of the convective flux in the v036 models. This is mostly significant in the HI region.

We present this effect in Fig.~\ref{fig:conv_struct}. The two panels show the vicinity of the two dynamically unstable regions: the left-hand side is the HeII zone, and the right-hand side is the HI zone. The left vertical axis is the flux, while on the right vertical axis, we present the nondimensional super-adiabatic gradient $Y$ with a green dotted curve, and the horizontal axis is the radial coordinate inside the star.  The real enthalpy flux, $F_c$, is denoted by the blue line. We separate the different regions described in the previous section by vertical dashed black lines. We also show the different terms of the convective energy flux: the temperature component $F_T=\overline{\rho}\,\overline{c_p} \widetilde{T^{\prime\prime}U_r^{\prime\prime}}$ is presented by an orange curve, the pressure term $F_p=\overline{\rho}\overline{(1-\delta)/{\rho}}\widetilde{p^{\prime\prime} U_r^{\prime\prime}}$ is given by the red solid line, and the molecular weight term, $F_\mu=\overline{\rho}\,\overline{(\partial h/\partial \mu)_{T,p}} \widetilde{\mu^{\prime\prime} U_r^{\prime\prime}}$ is denoted by the dash-dotted yellow line. This last term is calculated only approximately in our study, and the sum of the aforementioned terms (black dashed curve) is not fully equivalent with $F_c$. Lastly, the purple curve denotes the turbulent kinetic energy flux defined by \citep{Viallet2013} $F_{kin} = \overline{\rho} \widetilde{e_{t,0}^{\prime\prime}U_r^{\prime\prime}}$. This term  does not take part in the enthalpy transport, and it is sometimes neglected \citep[see, e.g., parameter set A and B in][]{Paxton2019}, but as we can see in Fig.~\ref{fig:conv_struct}, it is more significant than $F_p$. This flux term appears near the edges of the total convective zone (i.e., not in the inter-zone counter-gradient layer).  Interestingly in the top of the HI zone, $F_{\rm kin}$ becomes positive, which phenomenon is not present in solar and red giant convection, but was found earlier in A-type stars, which have thin convection zones similarly to the models presented here \citep{Kupka2009}.

\subsection{Structure and anisotropy of the velocity field}
\label{sec:veolicity}

\begin{figure}[htb!]
    \centering
    \includegraphics[width=\columnwidth]{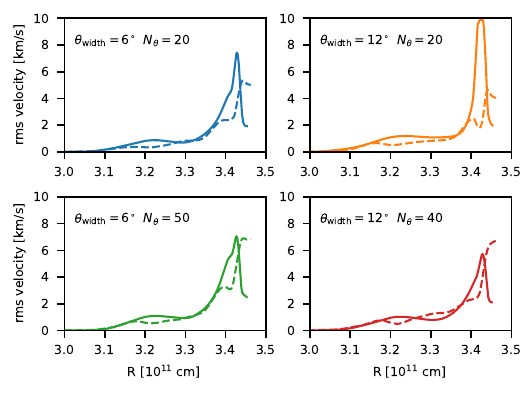}
    \caption{Velocity fluctuations in different models. The solid line denotes the radial direction, and the dashed lines denote the horizontal direction. Top left: {\tt v036\_6\_150x20}; top right: {\tt v036\_12\_150x20}; bottom left: {\tt v036\_6\_150x50}; bottom right: {\tt v036\_12\_150x40}. }
    \label{fig:rms_velocities}
\end{figure}

Now, we turn our attention to the horizontal fluctuations of the velocity fields. In Fig.~\ref{fig:rms_velocities}. we present the velocity fluctuations ($u_{r,\rm rms}$ with solid and $u_{\theta,\rm rms}$ with dashed line) versus the radius of four different models of v036: \verb|v036_6_150x20| (top left), \verb|v036_12_150x20| (top right), \verb|v036_6_150x50| (bottom left) and \verb|v036_12_150x40| (bottom right). These four models cover the observed trends. The rms velocities show two peaks, a smaller one in the HeII ionization zone at $R\sim 3.2\times 10^{11}$ cm, and a larger one in the overlapping HeI and HI ionization zone at $R\sim 3.4 \times 10^{11}$ cm. The slope of the radial velocity shows a hump at roughly the middle of the HeI region. The figure shows the presence of the Rayleigh-Bénard cells: we can see that in the middle of the convection zones, one has higher radial than horizontal velocities.

This structure also mirrors the structure seen in Fig.~\ref{fig:conv_struct}. The two dynamically unstable regions generate their own convective shells. Still, in the inter-zonal region, these layers overshoots toward each other, establishing a loose connection between the two zones.

\begin{figure}[htb!]
    \centering
    \includegraphics[width=\columnwidth]{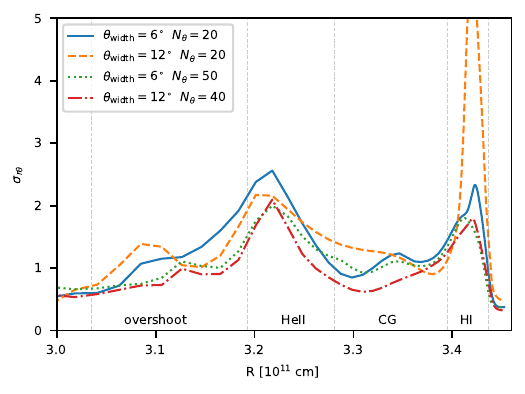}
    \caption{Anisotropy factors, $\sigma_{r\theta}$, of four different v036 models vs. the radial coordinate. The models are the reference model, {\tt v036\_6\_150\_20} (blue solid line); {\tt v036\_12\_150x20} (orange dashed line); {\tt v036\_6\_150x50} (green dotted line); {\tt v036\_12\_150x40} (red dash-dotted line).  }
    \label{fig:anisotropy}
\end{figure}

The anisotropy of the flow follows the same pattern in every model. Generally, the convective zones are anisotropic with $\sigma_{r\theta} \sim 2$, while in the counter-gradient and overshoot layers, it is almost isotropic. Mostly, we can say that the counter-gradient and overshooting layers show isotropic decaying of the granulation structure, as shown 
in Fig.~\ref{fig:anisotropy}. In the case of bad resolution $\Delta\theta > 0.2 ^\circ$, we experience a drop in the horizontal velocities and an increase in the radial velocities. In the case of resolutions with $\Delta\theta >1^\circ$, numerical viscosity is so large that the overall convective velocities drop.
Compared to the reference case of $\theta_{\rm width} =6^\circ$, $N_\theta=20$, increasing the size of the computational domain and increasing the resolution has a similar effect on $u_{\theta,\rm rms}$: it is increased, especially near the surface.

\subsection{General properties of the flow structure}
\label{sec:numbers}
\begin{figure}[htb!]
    \centering
    \includegraphics[width=\columnwidth]{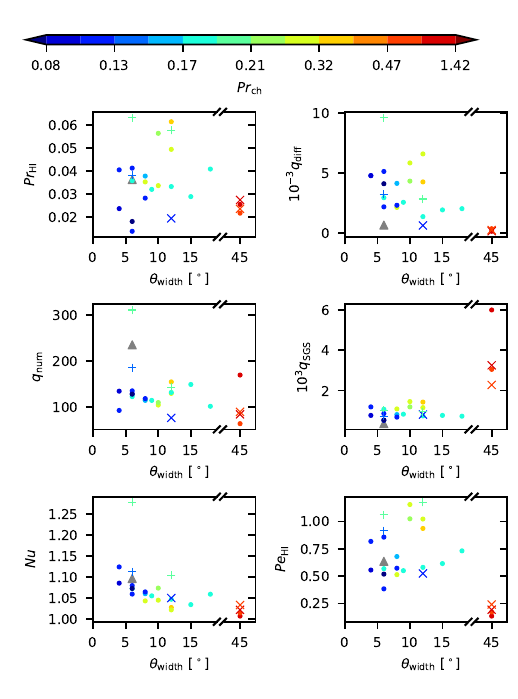}
    \caption{Dimensionless properties of the simulations vs. the angle width. The color-coding denotes the characteristic Prandtl numbers.  The symbols are the same as in Figure~\ref{fig:conv-prop}. The corresponding data can be found in Table~\ref{tab:model_properties}. Upper left: Prandtl number in the middle of the HI zone;
    Upper right: Ratio of numerical to SGS diffusivity; Middle left: Ratio of numerical viscosity to SGS viscosity; Middle right: Ratio of the SGS kinetic energy to the  turbulent kinetic energy of the resolved scales; Lower left: Maximum Nusselt number; Lower right: Péclet number in the middle of the HI zone.}
    \label{fig:conv-numbers}
\end{figure}

The dimensionless numbers describing the properties of the flow are presented against the $Pr_{\rm ch}$ and $\theta_{\rm width}$ in Fig.~\ref{fig:conv-numbers}, and are listed in Table~\ref{tab:model_properties}. We also present a list of grid aspect ratios, horizontal Reynolds numbers, and effective Prandtl and Péclet numbers at different radial positions in Appendix~\ref{appendix:numbers}. From the characteristic Prandtl and Reynolds number, one can conclude that the resolution presented here is too low to show actual turbulence in the simulations. The effective Prandtl numbers in the middle of the HI zone ($Pr_{\rm HI}$) are in the same range for every model: $Pr_{\rm HI} \in [0.02;0.06]$ (top left panel). This is systematically lower than $Pr_\text{ch}$ by a factor of 2--10, which is a combined effect of our arbitrarily chosen $\chi_\text{ch}$,  the fact that viscosity $\nu\propto \boldsymbol{U} $, and that $|\boldsymbol U | < U_{\rm ch}$.\footnote{The relatively low advective velocities are a consequence of inefficient convection in our models. This also decreases Prandtl and Péclet numbers as radiative diffusion can largely exceed numerical diffusion. This scenario is completely different from the convection at the bottom of the solar convection zone or above a convective
core of intermediate to high-mass stars.} This can be seen especially in the case of the very low-resolution models, with $\theta_\text{width}=45^\circ$. The ratio of numerical to SGS diffusion is in the same range, weakly anti-correlating with the opening angle (top right panel), and similar behavior can be observed in the case of the ratio of numerical to SGS viscosities (middle left panel). Both quantities are extremely large ($q_{\rm diff} \sim \mathcal{O}(1000)$; $q_{\rm num} \sim \mathcal{O}(100)$) which also indicates very large numerical damping, which easily outweighs the SGS model. Therefore, the ratio of SGS energy to total kinetic energy is low: $q_{\rm SGS} \approx 10^{-3}$ (middle right panel), and it is slightly increased in the bad resolution models, which is expected.

In the v036 models, the convective flux is inefficient and strongly depends on the opening angle (bottom left panel); meanwhile, Péclet numbers are similar in the different models: $Pe \approx 0.75$, except for the low-resolution model. The v046 models have larger convective fluxes ($Nu>1.1$), but still smaller than one would expect from 1D models \citep{KovacsGB2023}.

It can be said altogether that the numerical properties and Nusselt number (similarly to $u_{\rm rms}^{\rm tot}$, see Fig.~\ref{fig:conv-prop}) anti-correlate with the opening angle of the models. At the same time, the other dimensionless quantities do not depend significantly on them. The resolution also has a much smaller effect, and it decreases $Pr_{\rm HI}$ (which is expected) while increasing $u_{\rm rms}^{\rm tot}$.  Meanwhile $q_{\rm diff}\gg 1$ and $q_{\rm num}\gg 1$ meaning that even though $q_\text{SGS} \ll 1$, this means in this case that the velocity is so much damped by the numerics on the grid scale, that the SGS model is just a second-order problem next to it. This means that $q_\text{SGS}$ cannot be used to determine the quality of the resolution.

Comparing these results with the calculations of v046, we can see that as it has a lower effective temperature, it also shows the same trends with the opening angle. Meanwhile, having increased  (\verb|v036_6_200x20|) or decreased vertical resolution does not produce significantly different results.

\subsection{Length scales of the turbulent convection}
\label{sec:lengthscales}

\begin{figure*}[htb!]
    \centering
    \includegraphics[width=\textwidth]{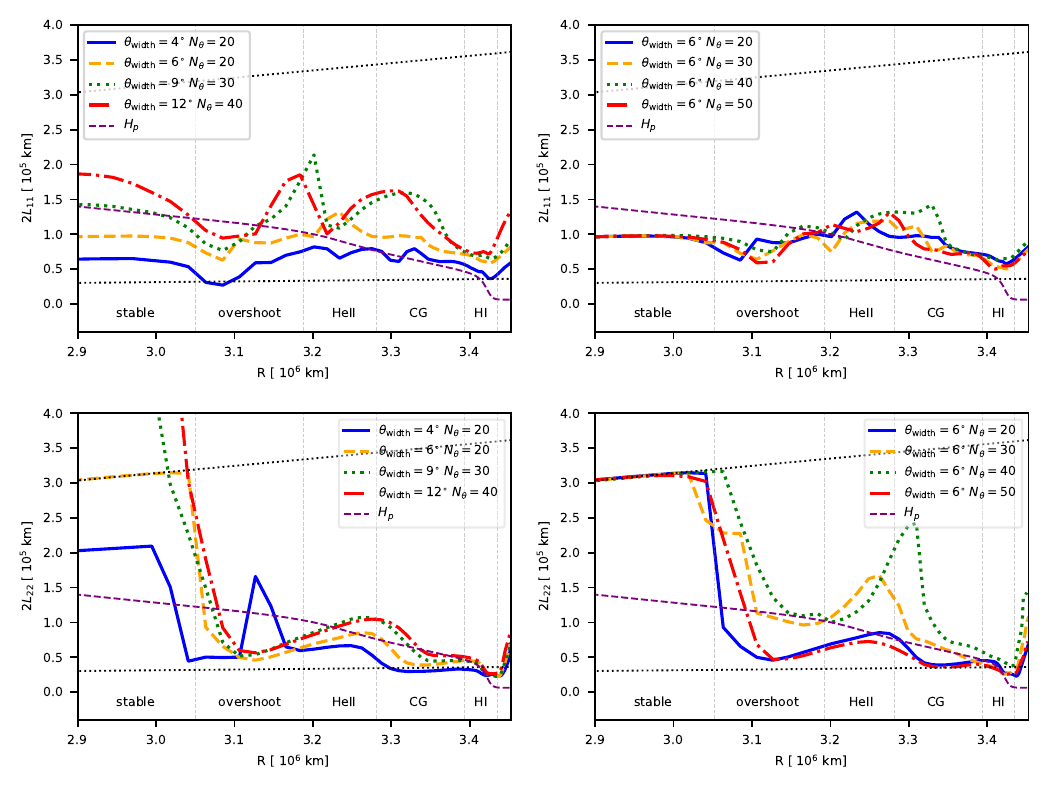}
    \caption{ Longitudinal (top panels) and transversal (bottom panels) integral length scales of different realizations as a function of the radial coordinate. Left  panels: Comparison of different $\theta_{\rm width}$ models; Right panels: Comparison of $\theta_{\rm width} = 6^\circ$ models with different resolutions. The vertical dashed lines separate the different regions of the convection zones: the dynamically stable zone, the overshooting region, the convection zone corresponding to the HeII ionization layer, the counter gradient zone that connects the HeI ionization and the HeII ionization layers, and the convective zone belonging to the overlapping HeI and HI ionization. The thin dotted lines denote the horizontal length of a cell (bottom) and the total horizontal domain in the $6^\circ$ wide model (top). }
    \label{fig:lenghtscales}
\end{figure*}

As shown in Fig.~\ref{fig:vorticity} and the supplementary animation, the horizontally larger models do not contain more vorticity structures in the HeII region; instead, the size of the structures is increased.
To further delve into this problem, we calculated the longitudinal and transversal integral length scales of the velocity field, as they give us information about the size of the largest eddies. In this section, we refer to the diameters of the eddies as $2L_{11}$ and $2L_{22}$. We note that we choose the reference models to $\Delta\theta = 0.3^\circ$ and $\theta_{\rm witdh}=6^\circ$, that was used in \citet{SPHERLSII}.

We present the results of seven chosen models in Fig.~\ref{fig:lenghtscales} to differentiate between the effects of different opening angles and resolutions. This figure shows $2L_{11}$ and $2L_{22}$ values in the top and bottom panels, respectively.
We chose the models for the presentation to cover the problem of resolution and opening angles. Hence, we compare the reference models to different $\theta_{\rm width}$ runs in the left-hand side panels, while the effects of resolution change compared to the reference model are presented in the right-hand side panels.

In Figure~\ref{fig:lenghtscales} top left panel we compare the $2L_{11}$ values of the model \verb|v036_4_150x20| (blue solid line),\verb|v036_6_150x20| (yellow dashed line),\verb|v036_9_150x30| (green dotted line) and \verb|v036_12_150x40| (red dash-dotted line). The horizontal axis is the radial coordinate, and we labeled the different subregions of the convective zone. We plotted the pressure scale height with a thinner purple dashed line for comparison. The resolution of the $\Delta\theta=0.3^\circ$ models is shown with a black dotted line at the bottom of the figure, while the total domain size of the $\theta_{\rm width}=6^\circ$ model in the different depths is denoted by the top black dotted line. We can see that the horizontal velocities, given a large enough domain, create eddies at the bottom and top of the HeII zone around $2/3$ of the reference ($\theta_{\rm width}=6^\circ$) models. In other words, the HeII zone's boundary regions produce eddies with sizes correlating with the computational domain size.

In the meantime, there is no significant change in the longitudinal integral length scales when we increase the resolution, as it can be seen in the top right panel of Fig.~\ref{fig:lenghtscales}
Here, the reference model (\verb|v036_6_150x20|) is the blue solid line, \verb|v036_6_150x30| is denoted by the yellow dashed line, a green dotted line is the \verb|v036_6_150x40| and red dash-dotted line presents the higher resolution \verb|v036_6_150x50|. As one can see, there are no significant differences between the longitudinal integral length scales of these models.

The angle width effect is present, albeit with much less efficiency, when we compare the transversal lengths scale in the bottom left panel of Fig.~\ref{fig:lenghtscales}; the models are the same as those in the top left panel. We note that in homogenous isotropic turbulence \citep{Pope-konyv}, $L_{22} = L_{11}/2$. Comparing the left-hand side panels to each other, and also with the anisotropies in Fig.~\ref{fig:anisotropy}, one can deduce that the  flow is indeed almost isotropic in the counter-gradient and overshooting regions. Corresponding to the larger anisotropy in the dynamically unstable regions, the $2L_{22}$ values are also larger. 

The increased resolution did not affect the $2L_{22}$ values, similarly to the case of $2L_{11}$. On the other hand, we detected a strange anomaly in the case of \verb|v036_6_150x30| and \verb|v036_6_150x40|, in which the transversal integral length scales are enhanced at the top boundary of the HeII zone. We show this phenomenon in the bottom right panel of Fig.~\ref{fig:lenghtscales}.  This behavior is puzzling and will need further investigation.

Another interesting feature of the length scales is the return to the horizontal size of the model in the stable region. As $|\boldsymbol U | \rightarrow 0$, the $R_{rr}(r_d)/R_{rr}(0)\rightarrow 1$, the transversal integral length scale quickly becomes the total horizontal width. Meanwhile, it seems that there is some remaining horizontal velocity in the deeper regions, which still shows a decaying auto-correlation. In the deeper 2D regions, $2L_{22}(r) = r\theta_{\rm width}$ while $2L_{11}(r) \approx r\theta_{\rm width}/3$. 

In summary, we can conclude that the diameter of the larger eddies is greatly dependent on the angular size of the modeling box until reaching $\theta_{\rm width}>9^\circ$. After that point, the effect diminishes. The larger resolution than the reference $\Delta\theta=0.3^\circ$ does not affect the larger eddies, and there is some remainder horizontal velocity in the dynamically stable region. 

\subsection{The behavior of the energy cascade}
\label{sec:energy}
An LES simulation aims to resolve the large, possibly anisotropic scales of motion, that is, the energy-containing ranges \citep{Pope-konyv} of the turbulent cascade. The filtering function that appears in the derivation of such a method \citep{LES_konyv} damps the energy cascade at the filtering scales (in our case, it is $l$, the grid scale length). The energy of these scales and (those that are below $l$) are modeled through the $e_{SGS}$ models. Having reached $\sim80\%$ of the kinetic energy in the resolved scales implies that the largest structures are resolved \citep{Pope-konyv}, but these will be severely damped by the numerical (and SGS) viscosities; therefore, a real turbulent energy cascade cannot be formed. For a good resolution, one would need at least $95-99\%$ in the resolved TKE, which also means that $q_\text{num} \ll 1$ and $q_{\rm SGS} \ll 1$. As we have seen in Section~\ref{sec:numbers}, our simulations are far from this.

 Nevertheless, numerical simulations of this scale can show an energy cascade that eventually have a power law subrange. This does not mean that we reached automatically the inertial subrange. In fact, the granulation pattern of stars (slow uprising and strong confined downflows) show a similar energy spectrum \citep{Nordlund1997}.

\begin{figure}[tb!]
    \centering
    \includegraphics[width=\columnwidth]{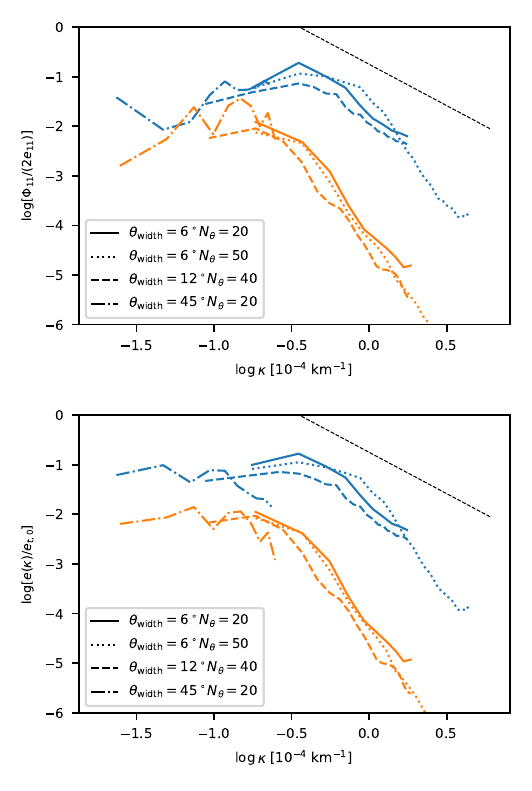}
    \caption{Logarithm of the  horizontal kinetic energy spectrum  normalized by $2e_{11}$ (top panel) and the total normalized turbulent kinetic energy spectrum  normalized by $e_{t,0}$ (bottom panel) vs. the logarithm of the horizontal wave-number in the HI zone (blue) and the HeII zone (orange) for different models: {\tt v036\_6\_150x20} (solid line); {\tt v036\_6\_150x40} (dotted line); {\tt v036\_12\_150x60} (dashed line); {\tt v036\_45\_150x20}  (dash-dotted line).  We show the Kolmogorov $-5/3$ slope as a dashed black line for comparison.   The low-resolution model  does not show the granulation energy cascade as it is damped already at the energy maximum. Simultaneously, the amount of total kinetic energy is lower in the HeII region, and $L_{11}$ and $L_{22}$ are larger. Thus, the spectrum has $\kappa_\text{max}$ at a lower wavenumber. }
    \label{fig:energy_spectrum}
\end{figure}

 To study the presence of granulation, we present the normalized horizontal kinetic energy spectra ($\Phi_{11}/(2e_{11})$) and normalized turbulent kinetic energy spectra ($e_t(\kappa)/e_{t,0}$) of four 
representative models in Fig.~\ref{fig:energy_spectrum} (top and bottom panels, respectively). We chose two different depths for the comparison: the middle of the HeII zone and the middle of the HI zone; the orange curves represent the former, while the blue curves denote the latter. The dashing type of the curves denotes the models: results of \verb|v036_6_150x20| is denoted by solid curves, \verb|v036_6_150x50| is presented by the dotted curve, dashed curve presents \verb|v036_15_150x40| and lastly the dash-dotted curve is the \verb|v036_45_150x20|. 

Taking  the strong damping into account, we accept a resolution (and horizontal size) if the structure of the energy spectrum is the following:
\begin{itemize}
    \item there is a driving range from the largest scales (smallest $\kappa$), in which the spectrum increases, reaching its maximum.
    \item This maximum ($\kappa_\text{max}$) corresponds to the largest scale eddies, and the energy is determined by the number of up and downflows that can be created.
    \item At $\kappa > \kappa_\text{max}$, $e(\kappa)$ starts decreasing, which range contains a part where $e(\kappa) \propto \kappa^{-5/3}$or even $\kappa^{-2}$ \citep{Nordlund1997}.
    \item This is followed by a steeper decay, where numerical/SGS viscosity takes over.
\end{itemize}

In the top panel of Fig.~\ref{fig:energy_spectrum}. we can see that in the bad-resolution models, the large-scale part of the spectrum is present, but it is sharply cut at the Nyquist-frequency, without reaching the decaying range. In addition, the spectrum fluctuates, and HI and HeII spectra are the same. The other models show that the real granulation spectra starts at wave numbers $\kappa_{\rm max} \sim3\times 10^{-3}$ km$^{-1}$. The  driving range start at lower wave numbers and are not  present in the $6^\circ$ wide models (lack of rising part). Otherwise, the larger resolution model resembles more closely the spectra presented by \citet{Nordlund1997}. 

If we look at the total turbulent kinetic energy spectrum in the bottom panel of Fig.~\ref{fig:energy_spectrum}, we can see that the vertical kinetic energy 
has different spectra in the two different regions. The resolution features are similarly present, though we can see that the dissipation range of the HI zone is not that strong with models $\Delta\theta = 0.3$. 
Meanwhile, in the HeII zone, {$\kappa_{\rm max} \sim 1.7\times10^{-3}$ km$^{-1}$}, that is, motions here are on a larger scale, which can be explained by the larger pressure scale height in this region.

\section{Discussion}
\label{sec:discussion}
We organize our discussion around the three main results: (1) the convective zone structure, (2) the domain size and resolution effects, and (3) the optimal size of RR Lyrae models based on our results.

\subsection{The consequences of the convective zone structure}

As we have seen in Sections \ref{sec:conv_struct} and \ref{sec:components}, the convective zone structure differs from the 1D picture, such as provided by the hydro-codes of \cite{Yecko1998} or  \citet{lengyel}. Recently, these two codes were compared by \citet{KovacsGB2023}. Primarily,  the two convectively unstable regions at the first hydrogen ionization zone and the second helium ionization zone are not completely separated in the static case \citep[as in the code of][]{lengyel}, but they are connected through decaying eddies, in a counter-gradient region.

The presence of counter-gradient layers is commonly known in astrophysics \citep{kupka}, but their actual presence in a flow can depend on various parameters. For instance, \citet{Kapyla2019} found a correlation between the extent of this zone and the effective Prandtl number in solar-like convection simulations.  Meanwhile, we do not find a CG region at the bottom of the HeII zone, not even in higher resolution models, suggesting that this region is very narrow, and we couldn't resolve it. In the RGB model of \citet{Viallet2013}, the transition between buoyancy driven and overshooting region is also very sharp. We conclude that this region would need an infeasibly large vertical resolution.

Nevertheless, the presence of CG layers  means that the general mixing length view currently employed in pulsation theory is not a good approximation in these regions. This approach assumes that $F_c \propto Y$ , that is, the convective flux is aligned with the super-adiabatic temperature gradient. The usage of this type of gradient-diffusion hypothesis creates various problems: overestimated convective velocities \citep{Wuchterl1998}, unphysical stratification \citep{Kupka2022,KovacsGB2023}, missing turbulent kinetic energy in the overshooting regions \citep{KovacsGB2023}. This theoretical problem has effects on the nonlinear behavior of pulsating stars \citep[see, e.g.,][]{lengyel2}, and probably is one of the main reasons for the reproduction difficulties of secondary light curve features \citep{Marconi2017rev}.

Our other main result is that due to the sharp stratification of the hydrogen partial ionization zone, the temperature flux cannot reproduce the whole convective flux: $\overline{\rho} \,\overline{ c_P} \widetilde{U^{\prime\prime}T^{\prime\prime}} \napprox \overline{\rho} \widetilde{U^{\prime\prime}h^{\prime\prime}}=F_c$ (see Fig.~\ref{fig:conv_struct}). Instead, we have latent heat transport, which originates from the ionization energy of the transported HII and HeII ions. This energy transport is significant, giving about half of the convective flux. In the meantime, \citet{Viallet2013} showed that the pressure term is nonnegligible if we use the entropy flux only \citep[see the derivation of][]{Kuhfuss1986}. This means that convection models that describe convective flux through solving temperature fluctuation \citep[e.g.,][]{Canuto1998} have to deal with this problem,  as well.  One way of doing this is to incorporate the fluctuations in the specific heat (in which ionization is usually incorporated)  into the definition of $F_c$. This method was investigated by \citet{Montgomery2004}, who found it to be a correction of $15\%$ in A-type stars as well as in DA and DB white dwarfs. We note that this is a specific feature of  sharp ionization fronts, and this problem probably does not occur in solar or red giant models. On the other hand, nonlocality of the convective-flux is a more serious problem in 1D models.

Another feature we found is the sign inversion of kinetic energy flux at the top of the HI zone. This phenomenon is not present in solar or red giant convection, instead it was reported in A-type star convection by \citet{Kupka2009}. \citet{Kupka2009} gives an explanation by pointing out that in their model the convective region is very shallow, causing heating from the bottom and cooling from the top acting on the same distance around one pressure scale height. Although they have presented this in a model without helium, our model  has lower density, and an even sharper temperature gradient. 

The structure of the velocity field is similar to the red giant and other models, similar to those in \citet{Viallet2013}, but the two convective zones can be separated as two maxima in the rms velocity profile 
 (Fig.~\ref{fig:rms_velocities}). In general, the large-scale convective motions are anisotropic (see Fig.~\ref{fig:anisotropy}), they become almost isotropic in the inter-zone counter-gradient layer and the overshooting region as well, which is supported by the behavior of integral length scales of this region ($L_{11} \approx 2 L_{22}$). The question of how this anisotropy influences the mode selection through pressure-like terms needs further study.

Altogether, the overall structure of the convective zones in static RR Lyrae models is composed of the two distinct dynamically unstable regions (also the engines of the pulsation), connected through a counter-gradient layer. This creates a very complicated coupling between these zones, which needs further studies for both the static and the dynamical models. The first step in unfolding this problem is determining the optimal model size and resolution to allow affordable computations with enough details. In the next section, we summarize and discuss our experiences with these models' size and resolution constraints.

\subsection{Effects of the domain size and resolution}

Nearly all the important quantities (size of the different structures in the convective zones, velocities, effectiveness of the convection, sizes of the eddies, anisotropy) showed some dependence on the horizontal size and resolution of the models.

In the case of the structure of the convective zone, the resolution effects become less important  at larger models (see  Sect.~\ref{sec:conv_struct} and Fig.~\ref{fig:conv-prop}). The horizontal width of the models had larger effects, and universally, we see that the mean convective velocity, $u_{\rm rms}^{\rm tot}$ decreases with the horizontal width of the models ($\theta_{\rm width}$) while the length scale of the overshooting zone, $\Lambda_{\rm OV}$ increases. This, together with the shape of the energy spectrum (i.e., $\kappa_\text{max} \sim 1.7 \times 10^{-3}$ km$^{-1}$) and the fact that the eddy sizes (Fig. \ref{fig:lenghtscales}) became unaffected by the angular sizes, show us that in the lower $\theta_{\rm width}$ models, the "box" of our simulation is too small, the largest two eddies interact with themselves through the periodic boundary condition. We note that with periodic boundary conditions, one always has at least two counter-rotating eddies \citep{conv-book}. 
As we presented in Sect.~\ref{sec:lengthscales}, the horizontal size severely affects the length scale of the overshooting ($\Lambda_{\rm OV}$) and nonzero turbulence ($\Lambda_t$) regions. The size of these regions are connected to the eddies formed in the HeII region. If $\theta_\text{width}$ is too small, the corresponding eddies will be small as well, resulting in a lower penetration depth.

The change in the other parameters can also be connected to the eddy sizes. 
The smaller eddies must increase the velocities to cover up the kinetic energy budget missing from the larger scales. This enhances convective flux in these models. 

Meanwhile, we have a stronger resolution problem in the horizontal direction, which increases numerical viscosity, outweighing the SGS model. This enhanced convection (especially in the HeII zone) will increase the Nusselt number. Due to the technical details of SPHERLS \citep{SPHERLS1}, the numerical viscosity is also dependent on the velocity, which will also be enhanced in this case (see Appendix~\ref{appendix:numerics}). Interestingly, we experience that the background pulsation is more effectively excited with smaller $ \theta_{\rm width}$, which is probably caused by the stronger instabilities of the flow. The supplementary animation also showcases the more violent nature (see also Fig.~\ref{fig:vorticity}) , which can be responsible for the increase of total turbulent kinetic energy of the \verb|v06_6_150x40| model seen in Fig.~\ref{fig:TKE-evol}.

The v046 models follow the same trends as the v036 models too, indicating that this might be a general feature of the RR Lyrae  LES models. On the other hand, it is important to note that despite being too small, the reference models that were also used in \citet{SPHERLSII,SPHERLSIII}, $\Delta\theta=0.3^\circ;\ \theta_{\rm width}=6^\circ$ models reproduced M3 RR Lyrae light curves reasonably well,  leading us to the conclusion that the effects are probably small compared to the classical pulsation instabilities themselves. This is due to the fact that classical pulsation itself is governed by the kappa-mechanism which needs good radial resolution, while 2D simulation even with low resolution can produce more accurate convective effects then 1D mixing length theory. These box sizes and resolutions are nevertheless inadequate to model strongly nonradial phenomena, such as  p-mode pulsation \citep[see, e.g., the resolution of the models of][]{Stein2001}. Meanwhile, studying the nonlinear, chaotic effects and the comparison with 1D models need the full view of which effects are numerical and which  are physical; our further studies will need larger $\theta_{\rm width}$ models.

The larger $L_{22}$ integral length scales of the models \verb|v036_6_150x30| and \verb|v036_6_150x40| in the bottom right panel of Fig.~\ref{fig:lenghtscales} remain a puzzle. The asymmetry of the resolution is an unlikely explanation as the  \verb|v036_6_150x50|  model has a similar $L_{22}$ profile, than \verb|v036_6_150x20|.  We hypothesize that some quasi-stationary structures are developed in these models. This can be a feature of the 2D modeling as well, since,  2D convection models are mostly less ergodic than the 3D ones \citep{Kupka2020}. Studying stochastic and quasi-static structures would be an interesting future project, as these can cause   modulations, for example, but are beyond of the scope of the present work.

The behavior of  horizontal structures that lower $L_{11}$ in the stable zone is probably a numerical artifact inherent to the periodic boundary condition. The original numerical fluctuations can normally cause a net drifting of the Rayleigh-Bénard cells \citep{conv-book}. Meanwhile, we note that the net velocities in the stable region are negligible ($u_{\theta,\rm rms} \ll 1$ km/s).

In the case of the velocity profiles, we see that increasing horizontal resolution and increasing the horizontal size of the models both decrease the anisotropy of the models. The effect in the bad resolution models is interesting: at around $\Delta\theta \approx 0.6^\circ$, we had enlarged radial velocities and decreased horizontal velocities at the convective boundary. In contrast, above $\Delta\theta > 1^\circ$, all kinetic energy terms are decreased. To understand this feature, we should look at the kinetic energy spectra of the models, as we can see in Fig.~\ref{fig:energy_spectrum}. The $\Delta\theta > 1^\circ$ model cannot catch the granulation, while the $\theta_{\rm width} = 12^\circ,\  \Delta\theta =0.3^\circ$ model shows similarities with the granulation spectra presented by \citet{Nordlund1997}. 

Another problem with the large models with bad resolution is the numerical damping.
With $q_\text{num} \gg 1$ and $q_\text{diff} \gg 1$, numerical damping acts on all lengthscales. Comparing our energy spectra to the one of \citet{Chan1996} or \citet{Porter2000}, we cannot reproduce the Kolmogorov cascade, and our energy spectra resembles the granulation pattern \citep{Nordlund1997}. Even with codes producing less damping, one needs at least a decade of wavenumbers, that are  affected by it \citep{Kritsuk2011}. In our models wavenumbers barely exceed one decade, and therefore all scales  are affected by the damping. In these circumstances, the effects of the SGS model cannot be studied.

The vertical resolution has smaller effects, as it has a constraint originating from the pulsation calculation itself{,  and therefore it is larger than the horizontal resolution}. The increased vertical resolution model (\verb|v036_6_200x20| with gray triangle in Fig.~\ref{fig:conv-prop}. and \ref{fig:conv-numbers}.) does not show significant differences from the standard 150 radial zone models. We ran some decreased vertical resolution models, with 122 radial zones (denoted by crosses in Fig.~\ref{fig:conv-prop}. and \ref{fig:conv-numbers}.). In these cases, having more horizontal cells had a larger effect than having less vertical cells. This mainly means that a model grid with resolution adequate for pulsation modeling \citep[see, for example, some description in][]{Paxton2019} is adequate to resolve the convective eddies, as well.

\subsection{Optimal size of 2D RR Lyrae models}

The largest obstacle is the calculation time of the models. As we previously presented in Fig.~\ref{fig:comput-times}, the SPHERLS code reaches API overhead above twelve cores. This number mostly depends on the vertical resolution, as horizontal parallelization is not implemented in the code due to calculating grid velocities \citep{SPHERLSII}. This means that roughly above 40 horizontal cells, we reach real-time computation speed. For a classical pulsator to reach maximum amplitude, it is needed to calculate hundreds or thousands of pulsation cycles \citep{bpf-beat2002}. To study mode selection, we need more \citep{bpf-drrlyr2004,Paxton2019}. Therefore, a real-time calculation with an RR Lyrae, for 1000 pulsation cycles could take almost one and a half years, which is not feasible.

Our other obvious constraints are the resolution and minimum horizontal sizes needed for a meaningful simulation. We found that the minimum angular size that is needed for the largest eddies in the second helium ionization zone is at least $9^{\circ}$--$10^{\circ}$, supported by the behavior of the longitudinal length scale and the appearance of the second  up and downflow structures. The longitudinal integral length scale can be also used for a resolution constraint \citep{Pope-konyv}, which is $2L_{11} \approx 1\times10^{10}$ cm which is around $\Delta\theta \approx 0.3^\circ$.  Meanwhile, the delay in the initial convective velocity growth (see Fig.~\ref{fig:TKE-evol}.) indicates, that $\Delta\theta < 0.2^\circ$ is needed for realistic driving. This way, we can conclude that \citet{SPHERLSII} had an adequate resolution to observe surface eddies, but for a meaningful comparison with 1D models, we need at least 1.5--2 times larger models with slightly better resolution. The $N_\theta=40$\ $\theta_{\rm width} = 12^\circ$ models are currently the maximum affordable models for our available hardware regarding domain size, while reaching stricter $\Delta\theta$ constraints one would need $N_\theta=45\ \theta_{\rm width}=9^\circ$. An optimistic choice of  $N_\theta=30\ \theta_{\rm width}=9^\circ$ models can be seen as a compromise solution for quicker calculations.

\section{Summary and conclusions}
\label{sec:conclusion}
We  performed a series of quasi-static envelope simulations with different angular widths and resolutions of two  M3 RR~Lyrae stars to determine the main features of the convection zone and to acquire an optimal setup for comparison with 1D model codes. The input parameters were previously derived by \citet{KovacsGB2023}. 

To achieve our goals, we analyzed the resulting models through a series of statistical methods used in turbulence theory \citep{Pope-konyv}. Under the assumption of the ergodic hypothesis, we calculated the fluctuating velocity (or root-mean-square difference) profiles $\sqrt{\widetilde{(U^{\prime\prime})^2}}$, the convective flux, and its components; we determined the size of the larger convective eddies by using longitudinal and transversal integral length scales. We also calculated the kinetic energy spectrum of the models in the two main convective zones.
Our results are the following:
\begin{enumerate}
    \item The convective zone in the studied RR Lyrae stars is complex. In the rms velocity profile, one can determine the two unstable zones at the overlapping first partial ionization regions of hydrogen and helium and the second ionization region of helium. We call these dynamically unstable regions HI and HeII, respectively. However, these two regions are connected in the velocity profiles by a counter-gradient layer.
    \item This inter-zonal counter-gradient layer and the bottom overshooting region are almost isotropic. The size of the overshooting region correlates with the angular size of the model, with a difference in the region's size with a full pressure scale length.
    \item We found positive kinetic energy flux in the top of the HI zone, which indicates very shallow convection, similar to A-type stars \citep{Kupka2009}.
    \item Due to the sharp stratification of the HI region, the traditional assumption that the convective flux can be described by temperature fluctuation alone does not hold. Approximately half of the energy in this zone is transported by the ionized matter as latent heat. This is because the stratification is so sharp and the ionization energies are so large that a small amount of displacement causes a large amount of energy difference. 
    \item The presence of a counter-gradient layer suggests that the relaxation of local diffusion based description of convection in 1D models should be a top priority in new radial pulsation codes.
    \item The smaller angular size models had larger convective velocities, more efficient convective transport, and more unstable flow structure. They had larger contamination of background pulsation, as well.
    \item This was caused by the enhanced convection that was an effect of the small box size, where eddies are not large enough, and the kinetic energy missing from the larger scales is compensated at the scales available in the model. The enhancements were negated at larger angular sizes. This process could also be followed by the change in the longitudinal and transversal integral length scales.
    \item Based on the correlation behavior \citep{ONeill2004} between the integral length scales ($L_{11}$ and $L_{22}$) and horizontal model sizes  ($\theta_{\rm width}$), as well as the appearance of the second up- and downflow structures in the HeII region, we  estimate a minimal necessary angular size of $\theta_{\rm width } \approx 9^\circ$.
    \item  Based on the condition of $\Delta_{filt} <= L_{11}$ \citep{LES_konyv,Pope-konyv}, the minimal necessary angular resolution was determined as $\Delta\theta =0.3^\circ$, while adequate driving of the convection needs at least $\Delta\theta \lesssim 0.2^\circ$, based on the initial velocity growth time. In addition, at  these resolutions the energy spectra resembles the formation of granules \citep{Nordlund1997}.
    \item The size difference in the eddies of the different convective zones  can also be followed in the horizontal and full kinetic energy spectra. The wavenumber of maximum energy ($\kappa_\text{max}$ is smaller in the HeII region, explained by the larger pressure scale heights in this region.
\end{enumerate}

The models presented here show a severe numerical damping on all scales, which can limit their usefulness. Based on the previous results of \citet{SPHERLSIII,SPHERLS4}, we can conclude that even with these problems, the optimal size and resolution presented here is enough to make conclusions about the large-scale convection and its representation in 1D models of classical variables.

Our current results of quasi-static RR Lyrae envelope convection pave the way toward a full-scale comparison with the 1D static envelope models, calculated with the Budapest-Florida Code \citep{Yecko1998,bpf-beat2002} and MESA RSP \citep{lengyel,Paxton2019}. The SPHERLS code already showed promising 2D results  \citep{SPHERLSII,SPHERLSIII,SPHERLS4}. In the following papers in our series, we provide a full analysis and comparison with the 1D models in addition to the static model structure differences covering the fully developed pulsation states.

This project helps  make our current knowledge more robust regarding the interaction between turbulent convection and pulsation. It also provides a valuable asset to improve our models in the future in their physical content and numerical realizations, exploiting the recently available hardware and software tools.

\begin{acknowledgements}
We are grateful to the referee, Dr. Friedrich Kupka for the careful and thorough reading of our manuscript, and whose detailed suggestions greatly improved the quality of this paper. We thank Robert Deupree and Chris Geroux for their generous help with the SPHERLS code.

This project has been supported by the `SeismoLab' KKP-137523 \'Elvonal grant, OTKA projects K-129249, SNN-147362, K-147131 and NN-129075, as well as the NKFIH excellence grant TKP2021-NKTA-64. The research was supported by the EK\"OP-24 University Excellence Scholarship Program of the Ministry for Culture and Innovation from the source of the National Research, Development and Innovation Fund, through the EK\"OP-24-4-I-ELTE-363 grant. This research was also supported by the International Space Science Institute (ISSI) in Bern/Beijing through ISSI/ISSI-BJ International Team project ID \#24-603 - “EXPANDING Universe” (EXploiting Precision AstroNomical Distance INdicators in the Gaia Universe).

On behalf of Project 'Hydrodynamical modeling of classical pulsating variables with SPHERLS' we are grateful for the usage of HUN-REN Cloud \citep[see][\url{https://science-cloud.hu/}]{H_der_2022} which helped us achieve the results published in this paper. 

\end{acknowledgements}
\bibliographystyle{aa} 
\bibliography{bibgraph}

\begin{appendix}
    \section{Governing equations of SPHERLS}
    \label{appendix:SPHERLS}
    
Here we present the governing equations of the SPHERLS code using a notation consistent with that one used in the paper. Further details are presented in the original instrument papers of \citet{SPHERLS1,SPHERLSII}. The main three governing equations are the conservation of mass, momentum, and energy:
\begin{align}
   {\rm D}_t \rho + \rho {\rm div}\, \boldsymbol{U} &= 0\text{,} \\
   {\rm D}_t \boldsymbol{U} &= - \frac{1}{\rho} {\rm grad\,} p - \frac{G M_r}{r^2} \boldsymbol{\hat{r}} + {\rm div\,}\boldsymbol{\underline{\tau}}^{\rm SGS} \text{,}\label{eq:velocity}\\
   {\rm D}_t e &=-\frac{p}{\rho} {\rm div\,} \boldsymbol{U} -\frac{1}{\rho}{\rm div\,} ( \boldsymbol{F}_r + \boldsymbol{F}_{\rm SGS} ) + D_{\rm SGS} \text{.}\label{eq:energy}
\end{align}
 Here ${\rm D}_t = \partial/\partial t + (\boldsymbol{U} - \boldsymbol{U}_0) \cdot \rm grad$ is the Stokes derivative, $\boldsymbol{U}$ is the velocity, $\boldsymbol{U}_0$ is the grid velocity, $p$ is the thermodynamic pressure, $\rho$ is the density, $G$ is the gravitational constant, $M_r$ is the total mass inside radius $r$, $\boldsymbol{\hat{r}}$ is the unity vector in radial direction, $\boldsymbol{\underline{\tau}}^{\rm SGS}$ is the subgrid scale turbulent stress tensor. In eq. (\ref{eq:energy}) $e$ denotes thermodynamic internal energy, $\boldsymbol{F}_{\rm SGS}$ the subgrid scale energy flux, and $D_{\rm SGS}$ describes energy transfer (dissipation) from subgrid scales to resolved scales. The radiative energy flux, $\boldsymbol{F}_r$, is described by the diffusion approximation,
\begin{equation}
    \boldsymbol{F}_r = -\frac{16\sigma T^3}{3 \rho \kappa_R} {\rm grad}\, T\text{,}
\end{equation}
 where $T$ is the temperature,   $\sigma$ is the Stefan-Boltzmann constant and $\kappa_R$ denotes the Rosseland mean opacity.  This approximation breaks down in the near-surface layers, where the atmosphere becomes optically thin, resulting in too hot layers, but classical pulsation is driven by the kappa mechanism deeper in the envelope. Since the kinematics and their coupling with convection matters more, 1D pulsation codes (and SPHERLS) have no aim to produce accurate atmospheres suitable for applications such as synthetic spectra calculations. Instead, all of these code produce bolometric light-curves which are then transformed to photometric bands via quasi-static atmosphere models \citep[see][]{Bono1994}. 
 
 The grid velocity is used to maintain radial resolution, especially in pulsating models, where due to the structural differences between the initial and full-amplitude models, a simple Eulerian grid wouldn't been able to maintain the resolution of the partial ionization zones, which drive the pulsation. This grid velocity is used to update the shell radii, while Lagrangian (mass) derivatives (i.e., $\partial/\partial r = 4\pi r^2 \overline{\rho} \partial/\partial M_r$) are used during the calculation in the radial direction. The grid velocity is derived from the net inflow and outflow of the total spherical shell \citep[see for details][]{SPHERLS1}:
\begin{equation}
    0 = \int \frac{\partial \rho}{\partial t} {\rm d}V + \oint \boldsymbol{U} \rho {\rm d}A - \oint \boldsymbol{U}_{0} \rho {\rm d}A \text{.}
\end{equation}

The horizontal boundary condition is periodic, so $\boldsymbol U_0$ only has a radial component, and the grid radius coordinates are updated by
\begin{equation}
    \frac{d r}{dt} = \boldsymbol{U}_0\boldsymbol{\hat{r}}\text{.}
\end{equation}

The subgrid scale effects are described by the Smagorinsky-model \citep{Smagorinsky1963}, and their formulas are the following \citep[and references therein]{SPHERLSII}:
\begin{align}
  & \nu_{\rm SGS} = \frac{{\rm C}^2 l^2}{\sqrt{2}}\left( {\rm grad}\, \boldsymbol{U} : [ {\rm grad}\, \boldsymbol U  + ({\rm grad}\, \boldsymbol U)^T]\right)^\frac{1}{2} \text{,} \\
  & \boldsymbol{\underline{\tau}}^{\rm SGS} = \rho \nu_{\rm SGS} \left[ {\rm grad}\, \boldsymbol U  + ({\rm grad}\, \boldsymbol U)^T- \frac{2}{3}\underline{\boldsymbol{I}}{\rm div}\,\boldsymbol U \right]\text{,}\\
  & \boldsymbol{F}_{\rm SGS} = -\frac{\nu_{\rm SGS}}{Pr_{\rm t}} {\rm grad}\, e\text{,}\\ 
  &  D_{\rm SGS} = \frac{{\rm d_t}}{L} \left(\frac{\nu_{\rm SGS}}{{\rm A_t} L}\right)^3\text{.}
\end{align}
 Here $\underline{\boldsymbol I}$ denotes identity matrix, $l$ is the cell scale length\footnote{$l$ is commonly called filter length scale and can be larger than the actual grid spacing. In that case we speak about subfilter scales, and dissipation is stronger on the grid scales, which can lead to numerically more stable calculations. However, the smallest scale where one has information in practice is $l$. \citep{Magnient2007}} ($l= \sqrt{\Delta r r \Delta \theta}$, with vertical grid spacing of $\Delta r$ and horizontal grid spacing of $\Delta \theta$), $L=3.75\times l$, ${\rm A_t}=0.117$, ${\rm d_t}=1.4$,$Pr_{\rm t}=0.7$, ${\rm C}=0.17$ are constants. The subgrid scale turbulent kinetic energy is related to the subgrid scale kinematic viscosity by the relation: $e_{\rm SGS} = [\nu_t /({\rm A_t} L)]^2$.

    \section{Numerical viscosity and artificial diffusion}
    \label{appendix:numerics}
Since there is no modeled molecular viscosity in the SPHERLS models (since it affects only scales that are much smaller than the model grid), to actually make our models comparable to real stars and other models, we need to derive some other measures instead.

In numerical calculations, the other source of dissipation next to molecular  (and SGS) viscosity  is numeric viscosity or  numerical damping, which is always unique to the numerical method. Its source is the truncation errors caused by finite differences.

 These truncation errors can be calculated using Taylor series \citep{Hirsch2007}. Since we deal with polar coordinates, one has to take into account the transform of the unit vector. Hence, we write the general Taylor series as:
\begin{multline}
\label{eq:Taylor}
  f(\boldsymbol x + \boldsymbol{\Delta x})= f(\boldsymbol{x}) + \boldsymbol{\Delta x} \cdot \grad f + \frac{1}{2}  \boldsymbol{\Delta x}^T \cdot (\grad \grad f )\cdot \boldsymbol{\Delta x} + \\  +\frac{1}{6}\boldsymbol{\Delta x} \cdot \left[\boldsymbol{\Delta x}^T \cdot (\grad\grad\grad f)\cdot \boldsymbol{\Delta x}\right] + \mathcal{O}(|\boldsymbol{\Delta x}|^4)\text{.}
\end{multline}

The displacement vectors are $\boldsymbol{\Delta x} = \Delta r \boldsymbol{\hat r}$ in the radial and $\boldsymbol{\Delta x} = r \Delta\theta  \boldsymbol{\hat \theta}$ in the horizontal direction (with $\boldsymbol{\hat \theta}$ as the covariant unit vector in the horizontal direction).

SPHERLS uses different discretization techniques in different situations. In general, it uses a staggered mesh, with fluxes defined at cell interfaces, internal variables defined at cell centers \citep{Deupree1977a,SPHERLS1}. Simple derivatives are calculated by the second-order accurate central method, while for the advection terms,  the donor-cell algorithm is used with weighted upwind and central schemes. The upwind scheme is the CIR scheme \citep{Courant1952}. The scheme is in 1D (with positive velocity direction),
\begin{equation}
\label{eq:donorcell} 
    \frac{{\rm d}f}{{\rm d}x} \equiv (1-\vartheta) \frac{f_{i+1} - f_{i-1}}{2\Delta x} + \vartheta \frac{f_{i} - f_{i-1}}{\Delta x} - \frac{\vartheta \Delta x}{2}\frac{d^2f}{dx^2}+\mathcal{O}{(h^3)}\text{,}
\end{equation}
where  $x$ denotes one of the coordinates, and the last term on the right-hand side is the second-order truncation error. The parameter $\vartheta$ is a weighting factor; its value is dependent on the velocities on the grid:\footnote{ $\min \vartheta = 0.2$ is a value used in SPHERLS, instead of the usual $0.1$, that was described in \citep{SPHERLS1}. This value is reported in \citet{Geroux-thesis}, as a trial-by-error solution.}
\begin{equation}
    \vartheta = \left\{ \begin{array}{ccc}
         1 & {\rm if}& |\boldsymbol{U}|/c_{\rm s} \ge 1, \\
         |\boldsymbol{U}|/c_{\rm s} &  {\rm if}& 1 > |\boldsymbol{U}|/c_{\rm s} \ge 0.2,\\
         0.2 & {\rm if}& |\boldsymbol{U}|/c_{\rm s} \le 0.2.
    \end{array} \right.
\end{equation}

The second-order truncation error is equivalent with an additional nonlinear diffusion term \citep{Pope-konyv}, which is in 2D Cartesian (assuming $\vartheta =1$):
\begin{multline}
\label{eq:diffusion}
    {\rm D}_t f = \partial_t f + U_x \partial_xf + U_y \partial_y f \\  \approx \partial_t f + U_x \delta_x f + U_y \delta_y f - \frac{U_x \Delta x}{2}\partial^2_xf-\frac{U_y\Delta y}{2}\partial_y^2f \\=  \partial_t f + U_x \delta_x f + U_y \delta_y f - {\rm div}\,\boldsymbol{F_{\rm num}}.
\end{multline}
Here $\delta$ denotes the finite difference operator describing the algorithm. The diffusivity is linear and $\chi^\text{num}_f = U\Delta x/2$ in 1D \citep{Pope-konyv}. However, in more than one dimension it becomes nonlinear and direction dependent. This nonlinearity is also enhanced by the unit vector transformations in polar coordinates.

Based on this, we first derive the second-order truncation errors in each direction for the velocity and energy advection terms. The velocity damping is called numerical viscosity. The damping terms can be derived from Eq.~(\ref{eq:Taylor}), and have the following forms:
\begin{multline}
\label{eq:Dampu_r}
    \mathcal{D}^U_r = \frac{(U_r-U_0)\Delta r}{2} \frac{\partial^2 U_r}{\partial r^2} + \frac{U_\theta \Delta \theta}{2}\frac{\partial^2U_\theta}{\partial r\partial\theta}\\ - \frac{U_\theta \Delta \theta}{2r} \frac{\partial U_\theta}{\partial \theta}+\frac{U_\theta \Delta \theta}{2}\frac{\partial U_r}{\partial r}-\frac{U_\theta U_r \Delta \theta}{2r}\text{,}
\end{multline}
\begin{multline}
\label{eq:Dampu_th}
    \mathcal{D}_\theta^U =\frac{(U_r-U_0)\Delta r}{2r}\frac{\partial^2 U_r}{\partial r\partial\theta} + \frac{U_\theta \Delta \theta}{2r}\frac{\partial^2 U_\theta}{\partial\theta^2} + \frac{(U_r-U_0)\Delta r}{2r^2}\frac{\partial U_r}{\partial \theta}\\ + \frac{U_\theta \Delta \theta}{2}\frac{\partial U_\theta}{\partial r} - \frac{(U_r-U_0)\Delta r}{2r}\frac{\partial U_\theta}{\partial r} + \frac{U_\theta^2 \Delta \theta}{2r}\text{,}
\end{multline}
\begin{equation}
\label{eq:dampe}
    \mathcal{D}_e = \frac{(U_r-U_0)\Delta r}{2}\frac{\partial^2 e}{\partial r^2}+\frac{U_\theta \Delta \theta}{2r}\left(\frac{\partial^2 e}{\partial\theta^2}+\frac{\partial e}{\partial r}\right)\text{.}
\end{equation}
Here $\mathcal{D}^U_r$, $\mathcal{D}^U_\theta$, and $\mathcal{D}_e$ are  the damping of $U_r$, $U_\theta$, and $e$, respectively. Assuming $\Delta r/r \ll 1$ and $r\Delta\theta/r \ll1$ we can neglect the last term in (\ref{eq:Dampu_r}) and the last two terms in (\ref{eq:Dampu_th}), arriving at
\begin{equation}
      D_t \boldsymbol U \approx \partial \boldsymbol{U} + \boldsymbol{U} \boldsymbol{\delta}\boldsymbol{U} - (\boldsymbol \nu_\text{num} \grad)(\grad \boldsymbol U)^T ,
\end{equation}
where the direction dependent numerical viscosity is
\begin{equation}
  \boldsymbol \nu_\text{num} = \left(  \frac{U_r \Delta r}{2} ,\quad \frac{U_\theta r \Delta \theta}{2}\right) .
\end{equation}
Then we take numerical viscosity as $\nu_\text{num} = |\boldsymbol \nu_\text{num}|$. Similarly one can derive from (\ref{eq:dampe}), that $\mathcal{D}_e =(\boldsymbol{\nu}_\text{num}\grad)\grad e \approx \chi^\text{num}_e {\rm div}\grad e$ with the numerical energy diffusivity $\chi^\text{num}_e = \nu_\text{num}$.  

Since the radiative heat conductivity is very large in our models, it is worth to study the numerical dissipation of the heat diffusion term ${\rm div} \boldsymbol{F}_r$. This is calculated by the second-order accurate scheme \citep{Hirsch2007},
\begin{multline}
\label{eq:radscheme}
    {\rm div} (K \grad T)_{i,j} = \\\frac{1}{\Delta x^2}\left[K_{i+1/2,j}(T_{i+1,j}-T_{i,j})-K_{i-1/2,j}(T_{i,j}-T_{i-1,j})\right] \\ + \frac{1}{\Delta y^2}\left[ K_{i,j+1/2}(T_{i,j+1}-T_{i,j})+K_{i,j-1/2}(T_{i,j}-T_{i,j-1})\right] \\+ \mathcal{O}(\Delta x^2 ,\Delta y^2),
\end{multline}
where $K=\chi \rho c_p$, and the radiative conductivity, and half-index values are $K_{i+1/2,j} = (K_{i+1}+K_i)/2$. We note that equation (\ref{eq:radscheme}) is in Cartesian coordinates for simplicity.  Truncation error here involves the third derivatives of the flux (i.e., fourth derivatives of temperature). Since this is a higher-order error than in the case of advection, we simplify the problem by considering $\boldsymbol F_r$ components as scalars. Therefore, the truncation error becomes:  
\begin{multline}
    \mathcal{D}_T = \frac{\Delta r^2}{6}\frac{\partial ^3 K (\grad T)_r}{\partial r^3}+\frac{\Delta\theta^2}{6r}\frac{\partial^3K(\grad T)_\theta}{\partial\theta^3}\\+\frac{\Delta\theta^2}{2}\frac{\partial^2K(\grad T)_\theta}{\partial r \partial \theta} -\frac{\Delta\theta^2}{6r}\frac{\partial K (\grad T)_\theta}{\partial\theta},
\end{multline}
from this we approximate numerical heat diffusivity as
    \begin{equation}
         \chi^{\rm num}_T \approx \frac{\mathcal{D}_T}{{\rm div}\,{\rm grad} T}\text{.}
    \end{equation}
This way, we provided all necessary numerical values. We note that these are usually only good approximations, as  truncation errors are highly nonlinear. 

On the other hand, to provide stability during shocks, SPHERLS also includes an artificial viscosity term \citep[][and references therein]{SPHERLS1} as an additional pressure term in equations (\ref{eq:velocity}) and (\ref{eq:energy}). However, we found this term is negligible in smooth regions.

\section{Numerical testing}
\label{appendix:tests}

\begin{figure*}[ht!]
    \centering
    \includegraphics[width=0.95\linewidth]{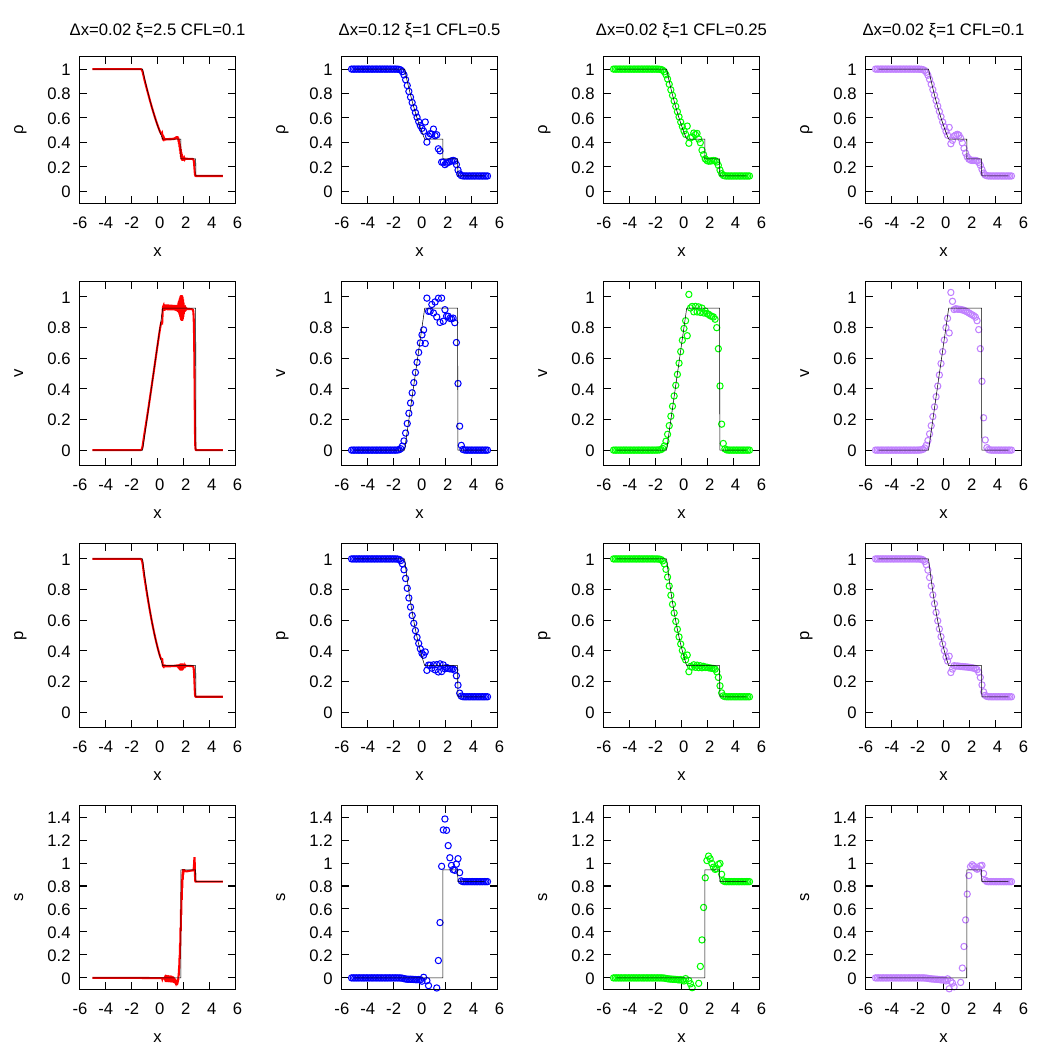}
    \caption{Results of the Sod shock-tube test for four scenarios: a higher resolution ($\Delta x=0.02$, first column), and three sparser grids ($\Delta x = 0.12$) with three different numbers: $0.5$, $0.25$, and $0.1$. The rows represent the density ($\rho$), velocity ($v$),  pressure ($p$), and specific entropy ($s$) (from top to bottom). }
    \label{fig:num_test}
\end{figure*}

The presence of the ionization front of RR Lyrae and other classical variables may question the usage of simple numerical methods that are used in SPHERLS. To investigate this problem, we  perform a numerical test of the donor-cell method against the Sod shock tube problem \citep{Sod1975} on an 1D Eulerian grid, solving Euler equations. The equation is given in the conservative form:
\begin{equation}
    \frac{\partial}{\partial t} \boldsymbol q + \frac{\partial}{\partial x}\boldsymbol F =0\text{,}
\end{equation}
where:
\begin{equation}
    \boldsymbol q = \left[\begin{matrix} \rho\\ \rho v\\ \frac{p}{\gamma-1}+\frac{1}{2}\rho v^2   \end{matrix}\right], \qquad \boldsymbol F =\left[\begin{array}{c}\rho v\\ \rho v^2 + p \\ \left(\frac{\gamma}{\gamma-1}p + \frac{1}{2}\rho v^2\right)v \end{array}\right].
\end{equation}
Here $\rho$ is the density, $v$ is the velocity, $p$ is the pressure, and $\gamma=1.4$ is the adiabatic constant. 
The equation in the test is solved explicitly by this scheme:\footnote{ The actual implementation of the donor-cell algorithm is more complex in SPHERLS, because it uses a staggered mesh with velocities defined at cell-borders, internal variables in the center of the cells, and (in the case of stellar problems) it uses implicit time steps. Nevertheless, this test is suitable to determine the general performance of this scheme}
\begin{equation}
    q^{*,n}_i = q^{n}_i-\frac{\Delta t}{\Delta x} (F_{i+\frac{1}{2}}-F_{i-\frac{1}{2}}),
\end{equation}
\begin{multline}
    q_i^{n+1} =  q^{*,n}_i\\ + \frac{\Delta t}{\Delta x} \xi\left[\left|q^{*,n}_{i+1}-q^{*,n}_{i}\right| \left(q^{*,n}_{i+1}-q^{*,n}_{i}\right) - \left|q^{*,n}_{i}-q^{*,n}_{i-1}\right| \left(q^{*,n}_{i}-q^{*,n}_{i-1}\right) \right].
\end{multline}
The second equation is the artificial viscosity in diffusion form \citep{Sod1975}. The half-centered flux terms are given by \ref{eq:donorcell}.

We present our result of a test\footnote{ The test case program is available on GitHub under MIT Licence at: \url{https://github.com/kovacsgb/upwindtest}} on the Sod shock tube problem in Figure~\ref{fig:num_test}. The Sod shock tube is characterized with a discontinuity with
\begin{equation}
    \rho_l =1, p_l=1,v_l = 0;\quad \rho_r =1/8, p_r=0.1, v_r=0.1\text{.}
\end{equation}

After the simulation is started, a shockwave starts to move to the right, which is followed by a slower contact discontinuity, while a rarefaction fan propagates to the left. This problem is a common test for numerical hydrocodes.

We performed one test with good resolution ($\Delta x=0.02$), and three tests on a sparser grid ($\Delta x =0.12$). All four tests were run to $t=1.4$ s, the time steps were determined by the Courant–Friedrichs–Lewy condition, with different CFL numbers (0.1 for the dense grid, and 0.5, 0.25, and 0.1 in the three sparse cases, respectively), such that
\begin{equation}
    CFL =\frac{ \max(|v_i| + c_i)\Delta t}{\Delta x}\text{.}
\end{equation}

In Fig.~\ref{fig:num_test}, we present the density, velocity, pressure, and entropy of the system (top to bottom). In the case of the dense grid, the scheme is highly unstable, as we needed to use a strong artificial viscosity with $\xi=2.5$ to achieve not disastrous result, but the oscillations are still visible, especially for the velocity and entropy. The sparse grids are more stable with $\xi=1$, we only experienced oscillations with $CFL=0.5$, while over and undershooting was present. The rarefaction fan is only slightly diffused, while the discontinuities show diffusion as well. Overall, this scheme shows similar features than the Ricthmyer scheme, but with stronger overshooting and oscillations.

Other numerical tests on an adaptive grid were made by \citet{Buchler1997} against the Noh \citep{Noh1987} and Sedov \citep{Sedov1959} problems, while in SPHERLS, the 2D Sedov-problem was investigated by \citet{SPHERLS1}.

It is worth to mention that in 1D pulsation codes \citep[e.g.,][]{Stellingwerf1975,Bono1994,lengyel} traditionally Lagrangian grids are used with simple upwind based flux terms. The sole exception is the adaptive code of the Budapest-Florida code \citep{Buchler1997,Yecko1998,bpf-beat2002}, where the \citet{vanLeer1977} scheme is used. Their numerical tests showed that the simple CIR scheme is inadequate to properly follow shock fronts \citep{Buchler1997}. This is an interesting result, considering the successes of 1D pulsation modeling \citep[see][Sec. 2. and references therein]{KovacsGB2023}. We propose a few answers to resolve this contradiction:
\begin{itemize}
    \item The ionization front is only investigated in the stellar envelope, while shock fronts are created above the photosphere, where the speed of sound drops. Therefore, the ionization front is mathematically not a discontinuity, and it can be resolved.
    \item All of these codes (including SPHERLS) use Lagrangian grids in which they resolve the ionization front, and its steep temperature gradient.
    \item The energy equation is solved implicitly, mostly using the \citet{Stellingwerf1975} method, or based on the method of \citet{Fraley1968}.
\end{itemize}
At a bottom line, SPHERLS can be considered as a pulsation hydrocode that is expanded into the horizontal direction. We can expect it to give better results for the convective processes, which happens naturally here, even if (especially horizontal) numerical dissipation is much stronger than in state-of-the-art LES/VLES/iLES codes.

\section{Flow and grid characteristics}
\label{appendix:numbers}

In this appendix, we provide the aspect ratios and Reynolds numbers of the models (Table~\ref{tab:appendix1}.), as well as the Prandtl and Péclet numbers (Table~\ref{tab:appendix2}.) in the two convective  regions and in the overshooting region. We note that our models have very inefficient convection, which means that radiative diffusion largely exceeds the numerical (and SGS) diffusion caused by calculation of the advective fluxes. This feature represents a flow  very different from those at the bottom of the solar convection zone or above a convective core of intermediate to high-mass stars, resulting in small Prandtl and Péclet numbers.

\begin{sidewaystable*}
\caption{ Grid aspect ratios defined by $r\Delta\theta/\Delta r$ and horizontal Reynolds numbers at different positions in the models.}\label{tab:appendix1}
\centering
\begin{tabular}{l||rr|rrr|r|rrr||rr|rrr|r|rrr} 
\hline
\multirow{3}{*}{Name} & \multicolumn{9}{|c||}{Aspect ratio} & \multicolumn{9}{|c}{Horizontal Reynolds number}\\
\cline{2-19}
& \multicolumn{2}{|c}{OV zone} & \multicolumn{3}{|c|}{HeII zone} &  \multirow{2}{*}{CG} & \multicolumn{3}{|c||}{HI zone}& \multicolumn{2}{|c}{OV zone} & \multicolumn{3}{|c|}{HeII zone} &  \multirow{2}{*}{CG} & \multicolumn{3}{|c}{HI zone}\\
\cline{2-6}\cline{8-15}\cline{17-19}
&\multicolumn{1}{c}{b} & \multicolumn{1}{c|}{m} & \multicolumn{1}{c}{b} & \multicolumn{1}{c}{m} & \multicolumn{1}{c|}{t} && \multicolumn{1}{c}{b} & \multicolumn{1}{c}{m} & \multicolumn{1}{c||}{t} & \multicolumn{1}{c}{b} & \multicolumn{1}{c|}{m} & \multicolumn{1}{c}{b} & \multicolumn{1}{c}{m} & \multicolumn{1}{c|}{t} && \multicolumn{1}{c}{b} & \multicolumn{1}{c}{m} & \multicolumn{1}{c}{t} \\
\hline\hline
\verb|v036_4_150x20|  & $ 0.54$ & $ 0.57$ & $  0.60$ & $ 0.7$ & $ 0.9$ & $  1.3$ & $  2.3$ & $    6.0$ & $   23$ & $  8$ & $   11$ & $   10$ & $   10$ & $  8$ & $  10$ & $  9.7$ & $  7.5$ & $    8.0$\\
\verb|v036_4_150x30|  & $ 0.34$ & $ 0.37$ & $ 0.42$ & $  0.5$ & $ 0.6$ & $ 0.9$ & $  1.7$ & $  4.2$ & $   16$ & $   11$ & $   16$ & $   15$ & $   16$ & $   15$ & $   14$ & $   13\phantom{.0}$ & $   11\phantom{.0}$ & $   15\phantom{.0}$\\
\verb|v036_6_150x20|  & $ 0.73$ & $ 0.82$ & $ 0.95$ & $  1.1$ & $  1.4$ & $    2.0$ & $  3.9$ & $  9.4$ & $   32$ & $  8$ & $   10$ & $   11$ & $   12$ & $  9$ & $  9$ & $  8.1$ & $  6.8$ & $  7.5$\\
\verb|v036_6_150x30|  & $ 0.49$ & $ 0.54$ & $ 0.63$ & $ 0.8$ & $ 0.9$ & $  1.3$ & $  2.6$ & $  7.1$ & $   22$ & $   17$ & $   13$ & $   13$ & $   14$ & $   14$ & $   12$ & $   11\phantom{.0}$ & $  8.5$ & $   11\phantom{.0}$\\
\verb|v036_6_150x40|  & $ 0.35$ & $  0.40$ & $ 0.48$ & $ 0.6$ & $ 0.7$ & $    1.0$ & $  1.9$ & $    4.0$ & $   21$ & $   26$ & $   19$ & $   18$ & $   23$ & $   25$ & $   19$ & $   16\phantom{.0}$ & $   14\phantom{.0}$ & $   17\phantom{.0}$\\
\verb|v036_6_150x50|  & $ 0.28$ & $ 0.32$ & $ 0.38$ & $ 0.5$ & $ 0.6$ & $  0.8$ & $  1.5$ & $  3.6$ & $   14$ & $   31$ & $   17$ & $   28$ & $   26$ & $   31$ & $   21$ & $   18\phantom{.0}$ & $   14\phantom{.0}$ & $   20\phantom{.0}$\\
\verb|v036_6_201x20|  & $  1.3\phantom{0}$ & $  2.2\phantom{0}$ & $   14\phantom{.00}$ & $   11\phantom{.0}$ & $  7.8$ & $  4.9$ & $  2.7$ & $  2.5$ & $  5$ & $  2$ & $  10$ & $   12$ & $   14$ & $   11$ & $  9$ & $  9.3$ & $  9.7$ & $    7.0$\\
\verb|v036_8_150x20|  & $ 0.94$ & $  1.1\phantom{0}$ & $  1.30$ & $  1.5$ & $  1.9$ & $  2.7$ & $  5.3$ & $   12\phantom{.0}$ & $   47$ & $  9$ & $  7$ & $   10$ & $  8$ & $  10$ & $   12$ & $  6.5$ & $  6.3$ & $  6.2$\\
\verb|v036_8_150x30|  & $ 0.62$ & $ 0.71$ & $ 0.85$ & $    1.0$ & $  1.3$ & $  1.8$ & $  3.9$ & $   10\phantom{.0}$ & $   31$ & $   14$ & $   12$ & $   19$ & $   13$ & $   17$ & $   15$ & $  8.7$ & $  7.5$ & $   10\phantom{.0}$\\
\verb|v036_8_150x40|  & $ 0.45$ & $ 0.52$ & $ 0.63$ & $ 0.8$ & $ 0.9$ & $  1.4$ & $  2.9$ & $  7.3$ & $   24$ & $   24$ & $   14$ & $   23$ & $   20$ & $   19$ & $   16$ & $   10$ & $  8.9$ & $   13\phantom{.0}$\\
\verb|v036_9_150x30|  & $ 0.68$ & $ 0.78$ & $ 0.95$ & $  1.1$ & $  1.4$ & $    2.0$ & $  3.9$ & $  8.9$ & $   35$ & $   17$ & $   11$ & $   20$ & $   12$ & $   16$ & $   16$ & $  8.5$ & $  7.6$ & $  7.9$\\
\verb|v036_10_150x20| & $  1.10$ & $  1.3\phantom{0}$ & $  1.60$ & $  1.9$ & $  2.4$ & $  3.3$ & $  6.1$ & $   30\phantom{.0}$ & $   43$ & $  8$ & $  5$ & $  7.4$ & $   11$ & $   10$ & $   13$ & $  8.7$ & $  6.5$ & $  2.8$\\
\verb|v036_10_150x30| & $ 0.78$ & $ 0.89$ & $  1.10$ & $  1.3$ & $  1.6$ & $  2.2$ & $    4.0$ & $   11\phantom{.0}$ & $   31$ & $   18$ & $   11$ & $   16$ & $   11$ & $   15$ & $   15$ & $  8.5$ & $  6.8$ & $  7.2$\\
\verb|v036_12_122x60| & $ 0.46$ & $ 0.53$ & $ 0.62$ & $ 0.8$ & $ 1.0$ & $  1.4$ & $  2.7$ & $  5.3$ & $   15$ & $   14$ & $  8$ & $   18$ & $   17$ & $   20$ & $   16$ & $   12\phantom{.0}$ & $   14\phantom{.0}$ & $   14\phantom{.0}$\\
\verb|v036_12_150x20| & $  1.30$ & $  1.5\phantom{0}$ & $  1.90$ & $  2.3$ & $  2.8$ & $  4.2$ & $   12\phantom{.0}$ & $   94\phantom{.0}$ & $   49$ & $  8$ & $  4$ & $  8$ & $    7$ & $  9.8$ & $   15$ & $  6.9$ & $  4.1$ & $  5.6$\\
\verb|v036_12_150x30| & $ 0.84$ & $    1.0\phantom{0}$ & $  1.30$ & $  1.5$ & $  1.9$ & $  2.6$ & $  4.9$ & $   21\phantom{.0}$ & $   38$ & $   15$ & $  5.6$ & $   14$ & $   12$ & $   10$ & $   10$ & $  5.9$ & $  5.6$ & $  4.6$\\
\verb|v036_12_150x40| & $ 0.65$ & $ 0.77$ & $ 0.95$ & $  1.1$ & $  1.4$ & $    2.0$ & $  4.4$ & $   11\phantom{.0}$ & $   36$ & $   22$ & $   11$ & $   18$ & $   13$ & $   17$ & $   13$ & $  8.3$ & $    8.0$ & $   11\phantom{.0}$\\
\verb|v036_15_150x50| & $ 0.63$ & $ 0.75$ & $ 0.95$ & $  1.1$ & $  1.4$ & $    2.0$ & $    4.0$ & $  8.2$ & $   45$ & $   28$ & $   17$ & $   16$ & $   14$ & $   17$ & $   17$ & $  8.7$ & $   11\phantom{.0}$ & $   14\phantom{.0}$\\
\verb|v036_18_150x60| & $ 0.68$ & $ 0.78$ & $ 0.95$ & $  1.1$ & $  1.4$ & $    2.0$ & $  3.9$ & $  9.6$ & $   36$ & $   34$ & $   17$ & $   19$ & $   15$ & $   17$ & $   17$ & $  8.2$ & $  8.1$ & $   11\phantom{.0}$\\
\verb|v036_45_122x40| & $  2.4\phantom{0}$ & $  2.8\phantom{0}$ & $  3.5\phantom{0}$ & $  4.3$ & $  5.5$ & $  7.7$ & $   16\phantom{.0}$ & $   28\phantom{.0}$ & $   94$ & $  7$ & $  4$ & $  6$ & $  4$ & $  5$ & $  5$ & $  3.8$ & $  3.5$ & $  2.5$\\
\verb|v036_45_122x60| & $  1.7\phantom{0}$ & $    2.0\phantom{0}$ & $  2.5\phantom{0}$ & $    3.0$ & $  3.7$ & $  5.2$ & $   10\phantom{.0}$ & $   19\phantom{.0}$ & $   58$ & $  7$ & $  5$ & $  6$ & $  5$ & $  6$ & $  7$ & $  4.7$ & $  4.4$ & $  3.2$\\
\verb|v036_45_150x20| & $  4.9\phantom{0}$ & $  5.7\phantom{0}$ & $  7.2\phantom{0}$ & $  8.5$ & $   11\phantom{.0}$ & $   15\phantom{.0}$ & $   31\phantom{.0}$ & $   53\phantom{.0}$ & $ 550$ & $  4$ & $  3$ & $  4$ & $  3$ & $    3$ & $  4$ & $  3.3$ & $  3.4$ & $  3.0$\\
\verb|v036_45_150x60| & $ 0.72$ & $  1.4\phantom{0}$ & $  9.1\phantom{0}$ & $  4.9$ & $  3.5$ & $  5.1$ & $   11\phantom{.0}$ & $   23\phantom{.0}$ & $ 130$ & $   25$ & $   12$ & $    4$ & $  5$ & $  5$ & $  5$ & $  3.8$ & $  3.7$ & $  4.1$\\
\verb|v046_6_152x20|  & $ 0.71$ & $  0.80$ & $ 0.92$ & $  1.1$ & $  1.4$ & $  1.8$ & $  2.4$ & $  6.4$ & $   24$ & $   11$ & $  9$ & $   15$ & $   13$ & $   14$ & $   12$ & $   11\phantom{.0}$ & $  8.1$ & $   12\phantom{.0}$\\
\verb|v046_6_152x30|  & $ 0.46$ & $ 0.53$ & $ 0.64$ & $ 0.8$ & $ 1.0$ & $  1.3$ & $  2.5$ & $  7.6$ & $   21$ & $   19$ & $   13$ & $   21$ & $   19$ & $   21$ & $   17$ & $   10\phantom{.0}$ & $   11\phantom{.0}$ & $   13\phantom{.0}$\\
\verb|v046_12_152x40| & $ 0.66$ & $ 0.78$ & $ 0.97$ & $  1.2$ & $  1.5$ & $    2.0$ & $  3.8$ & $   12\phantom{.0}$ & $   27$ & $   22$ & $  9$ & $   11$ & $   11$ & $   12$ & $   12$ & $  9.7$ & $  8.7$ & $  9.1$\\
\hline
\end{tabular}
\tablefoot{ The letters b,m,t stands for bottom, middle, and top, respectively. The chosen positions are: bottom and middle of the overshooting (OV) region, the bottom, middle, and top of the second Helium ionization zone, the middle of the inter-zonal counter-gradient (CG) layer, and the bottom, middle, and top of the overlapping first Helium and Hydrogen partial ionization zone.}
\end{sidewaystable*}

\begin{sidewaystable*}

\caption{ Effective Prandtl and Péclet numbers at different positions in the models. 
}\label{tab:appendix2}
\centering

\begin{tabular}{l||cc|ccc|c|ccc||cc|ccc|c|ccc} 
\hline
\multirow{3}{*}{Name} & \multicolumn{9}{|c||}{Effective Prandtl number} & \multicolumn{9}{|c}{Effective Péclet number}\\
\cline{2-19}
& \multicolumn{2}{|c}{OV zone} & \multicolumn{3}{|c|}{HeII zone} &  \multirow{2}{*}{CG} & \multicolumn{3}{|c||}{HI zone}& \multicolumn{2}{|c}{OV zone} & \multicolumn{3}{|c|}{HeII zone} &  \multirow{2}{*}{CG} & \multicolumn{3}{|c}{HI zone}\\
\cline{2-6}\cline{8-15}\cline{17-19}
& b & m & b & m & t && b & m & t & b & m & b & m & t && b & m & t \\
\hline\hline

\verb|v036_4_150x20|  & $0.001$ & $0.002$ & $0.006$ & $0.010$ & $0.006$ & $0.003$ & $0.005$ & $0.040$ & $0.007$ & $ 0.010$ & $ 0.024$ & $ 0.06$ & $  0.10$ & $ 0.07$ & $ 0.041$ & $ 0.081$ & $ 0.81$ & $  0.10$\\
\verb|v036_4_150x30|  & $0.001$ & $0.003$ & $0.008$ & $0.007$ & $0.004$ & $0.002$ & $0.004$ & $0.023$ & $0.005$ & $ 0.002$ & $ 0.029$ & $ 0.09$ & $ 0.11$ & $ 0.06$ & $ 0.039$ & $ 0.078$ & $ 0.55$ & $ 0.12$\\
\verb|v036_6_150x20|  & $0.001$ & $0.003$ & $0.011$ & $0.009$ & $0.007$ & $0.003$ & $0.005$ & $0.036$ & $0.010$ & $ 0.004$ & $ 0.033$ & $ 0.14$ & $ 0.17$ & $  0.10$ & $ 0.047$ & $ 0.077$ & $ 0.56$ & $ 0.14$\\
\verb|v036_6_150x30|  & $0.001$ & $0.002$ & $0.010$ & $0.009$ & $0.006$ & $0.003$ & $0.005$ & $0.041$ & $0.007$ & $ 0.006$ & $ 0.031$ & $ 0.11$ & $ 0.15$ & $ 0.11$ & $ 0.044$ & $ 0.094$ & $ 0.85$ & $ 0.14$\\
\verb|v036_6_150x40|  & $0.001$ & $0.002$ & $0.017$ & $0.014$ & $0.008$ & $0.003$ & $0.003$ & $0.014$ & $0.008$ & $ 0.010$ & $ 0.034$ & $ 0.27$ & $ 0.26$ & $ 0.24$ & $ 0.085$ & $ 0.079$ & $ 0.38$ & $ 0.23$\\
\verb|v036_6_150x50|  & $0.001$ & $0.002$ & $0.013$ & $0.012$ & $0.006$ & $0.002$ & $0.003$ & $0.018$ & $0.005$ & $ 0.010$ & $ 0.025$ & $ 0.23$ & $ 0.20$ & $ 0.18$ & $ 0.051$ & $ 0.090$ & $ 0.51$ & $ 0.14$\\
\verb|v036_6_201x20|  & $0.002$ & $0.002$ & $0.010$ & $0.008$ & $0.007$ & $0.003$ & $0.007$ & $0.036$ & $0.014$ & $ 0.008$ & $ 0.040$ & $ 0.27$ & $ 0.24$ & $ 0.14$ & $ 0.054$ & $ 0.130$ & $ 0.63$ & $  0.15$\\
\verb|v036_8_150x20|  & $0.001$ & $0.003$ & $0.014$ & $0.011$ & $0.010$ & $0.004$ & $0.006$ & $0.035$ & $0.014$ & $ 0.007$ & $ 0.025$ & $  0.20$ & $ 0.15$ & $ 0.15$ & $ 0.081$ & $ 0.072$ & $ 0.51$ & $ 0.19$\\
\verb|v036_8_150x30|  & $0.001$ & $0.003$ & $0.015$ & $0.012$ & $0.010$ & $0.003$ & $0.005$ & $0.038$ & $0.009$ & $ 0.010$ & $ 0.041$ & $ 0.34$ & $ 0.19$ & $ 0.23$ & $ 0.082$ & $ 0.078$ & $ 0.68$ & $ 0.16$\\
\verb|v036_8_150x40|  & $0.001$ & $0.002$ & $0.015$ & $0.012$ & $0.008$ & $0.003$ & $0.004$ & $0.028$ & $0.008$ & $ 0.015$ & $ 0.022$ & $ 0.35$ & $ 0.25$ & $ 0.19$ & $ 0.066$ & $ 0.077$ & $ 0.57$ & $ 0.16$\\
\verb|v036_9_150x30|  & $0.001$ & $0.002$ & $0.016$ & $0.012$ & $0.010$ & $0.004$ & $0.005$ & $0.032$ & $0.009$ & $ 0.014$ & $ 0.027$ & $  0.40$ & $  0.20$ & $ 0.24$ & $0.100$ & $ 0.075$ & $ 0.55$ & $ 0.13$\\
\verb|v036_10_150x20| & $0.001$ & $0.002$ & $0.018$ & $0.014$ & $0.011$ & $0.004$ & $0.010$ & $0.033$ & $0.015$ & $ 0.011$ & $ 0.020$ & $ 0.26$ & $ 0.31$ & $ 0.23$ & $0.110$ & $ 0.170$ & $  1.20$ & $ 0.09$\\
\verb|v036_10_150x30| & $0.001$ & $0.003$ & $0.017$ & $0.013$ & $0.010$ & $0.004$ & $0.007$ & $0.056$ & $0.011$ & $ 0.020$ & $ 0.037$ & $ 0.38$ & $ 0.19$ & $ 0.24$ & $0.110$ & $  0.100$ & $    1.00$ & $ 0.14$\\
\verb|v036_12_122x60| & $0.001$ & $0.002$ & $0.009$ & $0.007$ & $0.006$ & $0.003$ & $0.004$ & $0.019$ & $0.009$ & $ 0.013$ & $ 0.021$ & $ 0.14$ & $ 0.12$ & $ 0.15$ & $0.060$ & $ 0.078$ & $ 0.52$ & $ 0.18$\\
\verb|v036_12_150x20| & $0.001$ & $0.002$ & $0.018$ & $0.013$ & $0.012$ & $0.005$ & $0.056$ & $0.061$ & $0.020$ & $ 0.015$ & $ 0.012$ & $ 0.28$ & $ 0.18$ & $ 0.23$ & $0.130$ & $ 0.960$ & $ 0.93$ & $ 0.21$\\
\verb|v036_12_150x30| & $0.001$ & $0.001$ & $0.015$ & $0.011$ & $0.010$ & $0.005$ & $0.008$ & $0.049$ & $0.007$ & $ 0.011$ & $ 0.006$ & $ 0.31$ & $ 0.23$ & $ 0.17$ & $0.084$ & $ 0.082$ & $    1.00$ & $ 0.08$\\
\verb|v036_12_150x40| & $0.001$ & $0.001$ & $0.014$ & $0.010$ & $0.010$ & $0.004$ & $0.005$ & $0.033$ & $0.009$ & $ 0.023$ & $ 0.019$ & $ 0.31$ & $ 0.19$ & $ 0.25$ & $0.082$ & $ 0.078$ & $ 0.58$ & $ 0.18$\\
\verb|v036_15_150x50| & $0.001$ & $0.001$ & $0.015$ & $0.012$ & $0.010$ & $0.004$ & $0.005$ & $0.029$ & $0.009$ & $ 0.033$ & $ 0.022$ & $ 0.29$ & $ 0.21$ & $ 0.25$ & $0.100$ & $ 0.081$ & $ 0.61$ & $ 0.21$\\
\verb|v036_18_150x60| & $0.001$ & $0.002$ & $0.013$ & $0.011$ & $0.009$ & $0.004$ & $0.006$ & $0.041$ & $0.011$ & $ 0.048$ & $ 0.048$ & $ 0.31$ & $ 0.22$ & $ 0.22$ & $0.110$ & $ 0.091$ & $ 0.73$ & $  0.20$\\
\verb|v036_45_122x40| & $0.001$ & $0.002$ & $0.022$ & $0.017$ & $0.020$ & $0.008$ & $0.007$ & $0.027$ & $0.022$ & $ 0.017$ & $ 0.015$ & $ 0.26$ & $ 0.17$ & $ 0.18$ & $0.071$ & $ 0.053$ & $ 0.19$ & $ 0.08$\\
\verb|v036_45_122x60| & $0.001$ & $0.006$ & $0.017$ & $0.011$ & $0.014$ & $0.007$ & $0.006$ & $0.023$ & $0.017$ & $ 0.014$ & $ 0.045$ & $  0.20$ & $ 0.13$ & $ 0.15$ & $ 0.074$ & $ 0.059$ & $ 0.24$ & $ 0.09$\\
\verb|v036_45_150x20| & $0.002$ & $0.007$ & $0.037$ & $0.032$ & $0.040$ & $0.020$ & $0.012$ & $0.025$ & $0.040$ & $ 0.016$ & $ 0.031$ & $  0.30$ & $ 0.26$ & $ 0.22$ & $0.120$ & $ 0.062$ & $ 0.13$ & $ 0.16$\\
\verb|v036_45_150x60| & $0.004$ & $0.003$ & $0.004$ & $0.006$ & $0.012$ & $0.005$ & $0.005$ & $0.021$ & $0.017$ & $ 0.084$ & $ 0.050$ & $ 0.03$ & $ 0.05$ & $ 0.12$ & $ 0.045$ & $ 0.039$ & $ 0.17$ & $ 0.12$\\
\verb|v046_6_152x20|  & $0.001$ & $0.003$ & $0.022$ & $0.019$ & $0.015$ & $0.009$ & $0.011$ & $0.063$ & $0.017$ & $ 0.009$ & $ 0.032$ & $ 0.46$ & $ 0.32$ & $ 0.33$ & $0.190$ & $ 0.240$ & $  1.10$ & $ 0.31$\\
\verb|v046_6_152x30|  & $0.001$ & $0.003$ & $0.024$ & $0.016$ & $0.011$ & $0.005$ & $0.006$ & $0.038$ & $0.012$ & $ 0.014$ & $ 0.041$ & $ 0.45$ & $ 0.32$ & $ 0.29$ & $0.120$ & $ 0.110$ & $ 0.91$ & $ 0.24$\\
\verb|v046_12_152x40| & $0.001$ & $0.003$ & $0.020$ & $0.015$ & $0.015$ & $0.006$ & $0.008$ & $0.057$ & $0.012$ & $ 0.037$ & $ 0.033$ & $ 0.27$ & $ 0.27$ & $ 0.28$ & $0.120$ & $ 0.140$ & $  1.20$ & $ 0.19$\\

\hline
\end{tabular}
\tablefoot{The letters b,m,t stands for bottom, middle, and top, respectively. The chosen positions are: bottom and middle of the overshooting (OV) region, the bottom, middle, and top of the second Helium ionization zone, the middle of the inter-zonal counter-gradient (CG) layer, and the bottom, middle, and top of the overlapping first Helium and Hydrogen partial ionization zone.}
\end{sidewaystable*}

\end{appendix}

\end{document}